\documentclass[a4paper,11pt]{article}
\pdfoutput=1 
\usepackage{jheppub} 
\usepackage[T1]{fontenc}
\usepackage[italian,english]{babel}
\usepackage{hyperref}
\hypersetup{
	colorlinks=true,
	linkcolor=[rgb]{0.0,0.42,0.89},
	filecolor=magenta,      
	urlcolor=[rgb]{0.0,0.42,0.89},
	citecolor=[rgb]{0.64,0.0,0.0},
}
\usepackage{ifpdf}
\usepackage{subfigure}
\usepackage{gensymb}
\usepackage{amssymb}
\usepackage{amsfonts}
\usepackage{slashed}
\usepackage{epsf}
\usepackage{rotating}
\usepackage{graphicx}
\usepackage{amsmath}
\usepackage{fancyhdr}
\usepackage{lineno}
\usepackage{babel, blindtext}
\usepackage{graphics}
\usepackage{color}
\usepackage{physics}
\usepackage{multirow}
\usepackage{pstricks}
\usepackage{xcolor}
\usepackage{multirow}
\usepackage{hhline}
\usepackage{cancel}
\usepackage{framed}
\usepackage{mathtools}
\usepackage{diagbox}
\usepackage{orcidlink}
\usepackage{float}
\usepackage{ulem} 
\usepackage[lmargin=1.7in,bmargin=0.2in,footskip=0.5in,total={6.9in,9.5in}]{geometry}

\newcommand{\nn}{\nonumber}
\newcommand{\lsim}{\mathrel{\mathop{\kern 0pt \rlap
			{\raise.2ex\hbox{$<$}}}
		\lower.9ex\hbox{\kern-.190em $\sim$}}}
\newcommand{\gsim}{\mathrel{\mathop{\kern 0pt \rlap
			{\raise.2ex\hbox{$>$}}}
		\lower.9ex\hbox{\kern-.190em $\sim$}}}

\newcommand{\be}{\begin{equation}}
	\newcommand{\ee}{\end{equation}}
\newcommand{\bea}{\begin{eqnarray}}
	\newcommand{\eea}{\end{eqnarray}}

\title{\boldmath Strongly electroweak phase transition with  $U(1)_{L_{\mu}-L_{\tau}}$ gauged non-zero hypercharge triplet}

\author[a]{Seong Chan Park,\orcidlink{0000-0003-0176-4355}}
\author[b,c]{Anirban Biswas,\orcidlink{0000-0002-3810-3326}}
\author[d,e]{Shilpa Jangid\, \orcidlink{0000-0001-6307-1234}}

\affiliation[a]{	Department of Physics,                                                                                           IPAP, Lab for Dark Universe, Yonsei University, 50 Yonsei-ro, Seodaemun-gu,
	Seoul 03722, Korea}
	\affiliation[b]{Department of Physics, Gaya College (A constituent unit of Magadh University, Bodh Gaya), Gaya 823001, India}
	\affiliation[c]{
		Department of Physics, School of Sciences and Humanities, SR University,Warangal 506371, India}
\affiliation[d]{
	Asia Pacific Center for Theoretical Physics (APCTP) - Headquarters San 31,
	Hyoja-dong, Nam-gu, Pohang 790-784, Korea}
	\affiliation[e]{Shiv Nadar IoE Deemed to be University, Gautam Buddha Nagar, Uttar Pradesh, 201314, India}

\emailAdd{sc.park@yonsei.ac.kr,anirban.biswas.sinp@gmail.com,shilpajangid123@gmail.com}
\abstract{ The extension of the Standard Model Higgs doublet with three non-zero hypercharge triplets is examined in this article. The triplets are charged under additional $U(1)_{L_{\mu}-L_{\tau}}$ symmetry. We investigate the stability of the  electroweak vacuum both at the tree-level, and the two-loop level. It is observed that the vacuum stability can be satisfied till Planck scale using two-loop $\beta$-functions. In contrary, due to the increase in positive effect from triplet degrees of freedom, the perturbative unitarity can be satisfied only till $10^{12}$ GeV. The parameter space allowed from the Planck scale stability is checked for the strongly electroweak first-order phase transition. The model satisfies the strongly first order phase transition for the triplet bare mass parameters till TeV scale due to the enough contribution to the cubic term from the triplet degrees of freedom. It is observed that this model foresee strongly first-order phase transition for all mass ranges until the degrees of freedom becomes heavy enough to decouple from the thermal bath. The benchmark points satisfying the strongly first-order phase transition are tested for the Gravitational wave signatures. The benchmark points allowed from Planck scale stability, strongly first-order phase transition also comes out to lie in the detectable frequency range of the LISA and BBO experiments.  }


\begin{document}
	
	\maketitle
	\flushbottom
	
	\section{Introduction} 
The Standard Model (SM) of particle physics appeared to be a complete theory after the Higgs boson was discovered in 2012 by CMS \cite{CMS:2012qbp} and ATLAS \cite{ATLAS:2012yve} experiments at the Large Hadron Collider (LHC). The Standard Model has been shown to have flaws by a number of theoretical and experimental evidences, despite being the most effective theory to date. Among them are the stability of the electroweak vacuum up to the Planck scale, non-zero neutrino mass, electroweak baryogenesis, and so on. The Standard Model was extended with either gauge group or extra degrees of freedom charged under the same gauge group as a result of these evidences. Real or complex singlet, doublet, or triplet representations of $SU(2)$ are the lowest scalar extensions that are feasible.\\  

It is still unclear how the Standard Model can account for the tiny non-zero neutrino mass, flavor mixing patterns, fermion mass spectra, and CP violation \cite{Xing:2020ijf}. To account for the tiny neutrino masses and large lepton flavor mixing, the findings of neutrino oscillation necessitate an extension of the Standard Model. In this circumstances, an additional gauge symmetry like $U(1)_{L_{\mu}-L_{\tau}}$, which plays a critical role in understanding the longstanding muon $(g-2)$ anomaly \cite{Biswas:2016yan,Banerjee:2018eaf,Ma:2001md}, can also provide an interesting mixing pattern among the different neutrino flavors. The neutrino mass and mixing pattern in the triplet extended SM has already been studied in detail in \cite{Chun:2003ej, Biswas:2017dxt}. The Type-(I+II) seesaw model extension to the Standard Model with an extra singlet has also been extensively discussed with $U(1)_{L_{\mu}-L_{\tau}}$ gauge symmetry in the context of muon $(g-2)$ anomaly, tiny neutrino mass, and lepton flavor mixing \cite{Zhou:2021vnf}.
Moreover, it is well known that the current neutrino oscillation data permits only
two independent elements of the neutrino mass matrix to be zero,
commonly known as the two-zero texture \cite{Fritzsch:2011qv,Xing:2002ta}.
The most important feature of two-zero texture is that in this case the neutrino
mass matrix depends only on five real parameters, and therefore has
very good predictability of the other unknown parameters like
the Dirac CP phase, mass hierarchy etc. In our previous work \cite{SJangid}, we have demonstrated that 
the two-zero pattern can be obtained naturally in Type-II seesaw model with
$U(1)_{L_{\mu}-L_{\tau}}$ gauge symmetry when there are at least three scalar triplets
with different $L_{\mu}-L_{\tau}$ charges. The benefit of these additional scalar triplets
are many folds. Firstly, they can help us to resolve the vacuum instability problem of
the Higgs potential. The Higgs vacuum stability in the context of
Type-II seesaw has been studied in\cite{Han:2022ssz,Haba:2016zbu,BhupalDev:2013xol}.
Therefore, it is fascinating to explore neutrino mass matrix with
two-zero texture together with vacuum stability and perturbative unitarity
of the Higgs potential.\\ 

Finally, as initially proposed by Sakharov \cite{Sakharov:1967dj}, simultaneous occurrence of the baryon number violation, C and CP violation, and out of equilibrium processes is required to explain the baryon asymmetry, i.e., the excess of matter over anti-matter in the Universe. A highly first-order electroweak phase transition is one feasible way to provide these conditions for electroweak baryogenesis. For Higgs boson mass greater than 80 GeV \cite{Kajantie:1996mn,Kajantie:1996qd,Csikor:1998eu}, the electroweak phase transition (EWPT) from the symmetric phase at high temperature to the broken phase is not strongly first-order in the case of SM, but rather a smooth crossover \cite{Aoki:1999fi}, which is not consistent at all with the measured Higgs boson mass of 125.5 GeV \cite{Kajantie:1995kf}. Accordingly, \cite{Morrissey:2012db}, the departure from out of thermal equilibrium is also insufficient. Even if the electroweak phase transition satisfies the out of equilibrium conditions, the C, CP, and baryon number violation provided by the electroweak interactions via sphalerons is insufficient to explain the observed asymmetry \cite{Gavela:1994dt}. This provides another justification for expanding the SM to include more scalar degrees of freedom in the $SU(2)$ representation, at the very least a singlet \cite{Espinosa:1993bs,Bandyopadhyay:2021ipw}, doublet \cite{Blinov:2015vma}, or triplet \cite{Bandyopadhyay:2021ipw}.
A detailed discussion of strongly first-order phase transition (FOPT) in Type-II seesaw has already been provided \cite{Zhou:2022mlz,Roy:2022gop,Jangid:2023lny}. According to \cite{Kuzmin:1985mm,Cohen:1990it,Cohen:1993nk, Quiros:1994dr,Rubakov:1996vz,Funakubo:1996dw, Trodden:1998ym,Bernreuther:2002uj,Morrissey:2012db,DiBari:2013rga}, the additional degrees of freedom of the scalar triplet contribute to the cubic term and effectively explain the baryon asymmetry of the Universe through electroweak baryogenesis. It has been discussed in \cite{Zhou:2022mlz} that the strongly electroweak phase transition can be satisfied in the mass range upto 550 GeV in the minimal Type-II seesaw scenario with one additional scalar triplet. The more scalar degrees of freedom contribute more to the cubic term and hence, make the phase transition dynamics stronger. Hence, it is interesting to check the effect on the phase transition dynamics from additional scalar degrees of freedom coming from three triplets.
The stochastic background of the Gravitational waves results due to  the electroweak phase transition from symmetric phase to broken phase via bubble nucleation. A new avenue for investigating physics beyond supersymmetric symmetry has recently been made possible by the observation of Gravitational waves using LIGO and VIRGO \cite{LIGOScientific:2016aoc,LIGOScientific:2017vwq, LIGOScientific:2018mvr,LIGOScientific:2020ibl}. A thorough discussion of this may be found in the references and Refs therein \cite{Mazumdar:2018dfl,Caprini:2019egz,Athron_2024}. Therefore, it is intriguing to look into the physics beyond supersymmetry in order to examine the detectable frequency ranges of the present and upcoming Gravitational wave observatories, including BBO \cite{Ungarelli:2005qb,Cutler:2005qq}, LIGO, and LISA \cite{LISA:2017pwj}.
Therefore, we examine three Type-II triplets to investigate Gravitational wave signals, the strongly first-order electroweak phase transition, and the stability of the Higgs vacuum. The following is the outline of the article.\\

Under $U(1)_{L_{\mu}-L_{\tau}}$ gauge symmetry, the scalar potential for the three Type-II triplets is addressed in \autoref{model}. In \autoref{stability}, the perturbative unitarity up to the Planck scale, two-loop stability, and tree-level stability criteria are covered in detail. In \autoref{stability}, the one-loop $\beta-$ functions for the quartic couplings are provided, and in \autoref{betaf1}, the specific two-loop equations for the running of quartic couplings are supplied. \autoref{ewpt} discusses the corresponding Gravitational wave signatures and tests the allowable benchmark points from the vacuum stability and perturbativity for the electroweak phase transition.

\section{Model setup}\label{model}
The Type-II seesaw consists of a scalar triplet with non-zero hypercharge in addition to the SM $SU(2)$ Higgs doublet. It is an interesting scenario in providing the non-zero neutrino mass via induced triplet vacuum expectation value (vev). The positive $m_{\Delta}^2$ quantity in the bare mass term for the Higgs triplet $m_{\Delta}^2 \rm Tr[\Delta^{\dagger}\Delta]$ does not induce vev for triplet through electroweak symmetry breaking. In contrast, the triplet vev is induced by an interaction term, i.e., $\mu H^T \Delta H$ as $v_{\Delta} \propto \mu v_h^2/m_{\Delta}^2$. The value of the triplet vev is constrained from the electroweak precision experiment, i.e., mainly from the $\rho$ parameter as $v_{\Delta} \leq 3.0$ GeV, and $\mu$ is considered to be smaller in comparison to the electroweak scale to achieve such smaller value of $v_{\Delta}$. An additional gauge symmetry $U(1)_{L_{\mu}-L_{\tau}}$, and its spontaneous symmetry breaking is crucial in not only explaining muon $(g-2)$ but also in generating neutrino masses and lepton flavor mixing.
This symmetry would forbid the interaction term $\mu H^T \Delta H$ at the tree-level which is crucial in inducing the triplet vev. Hence, the Higgs triplet vev can be generated by adding new particles to the theory either at the tree-level or at loop-level effect. The gauge invariant Yukawa Lagrangian for Type-II seesaw with $U(1)_{L_{\mu}-L_{\tau}}$ is written as: 
\bea\label{eq:2.1}
\mathcal{L} = -y_l^{\alpha} \overline{l_{\alpha L}}\alpha_R H -\frac{1}{2}y_{\Delta} (\overline{l_{eL}} \Delta i \sigma_2 l_{\tau L}^C + \overline{l_{\tau L}} \Delta i \sigma_2 l_{e L}^C) + h.c.,
\eea

If the $U(1)_{L_{\mu}-L_{\tau}}$ charge for the triplet field is non-zero, in that case, the mixing pattern will not improve in the Type-II seesaw Majorona neutrino mass matrix. Hence, in order to achieve, less than three independent zero entries in the Majorona neutrino mass mixing matrix, the minimal requirement is to consider three scalar triplets \cite{SJangid}. 

The full relevant gauge invariant scalar potential with three additional $SU(2)$ triplets to the $SU(2)$ Higgs doublet with $U(1)_{L_{\mu}-L_{\tau}}$ symmetry is written as follows;

\begin{align}\label{scalar}
	V  \ = \ &  m_{\Phi_1}^2 (\Phi_1^{\dagger}\Phi_1)+ \lambda_{\Phi1} (\Phi_1^{\dagger}\Phi_1)^2 +  m_{\Delta1}^2 Tr(\Delta_1^{\dagger}\Delta_1)+m_{\Delta2}^2 Tr(\Delta_2^{\dagger}\Delta_2)+m_{\Delta3}^2 Tr(\Delta_3^{\dagger}\Delta_3) +\lambda_{\Delta1}(Tr(\Delta_1^{\dagger}\Delta_1))^2
	\nn \\     &    + \lambda_{\Delta2}Tr(\Delta_1^{\dagger}\Delta_1)^2  +\lambda_{\Delta3}(Tr(\Delta_2^{\dagger}\Delta_2))^2   + \lambda_{\Delta4}Tr(\Delta_2^{\dagger}\Delta_2)^2 
	+\lambda_{\Delta5}(Tr(\Delta_3^{\dagger}\Delta_3))^2  + \lambda_{\Delta6}Tr(\Delta_3^{\dagger}\Delta_3)^2 	\nn \\  
	& 
	+\lambda_{\Phi_1 \Delta1}(\Phi_1^{\dagger}\Phi_1)Tr(\Delta_1^{\dagger}\Delta_1)
	+ \lambda_{\Phi_1 \Delta2} \Phi_1^{\dagger}\Delta_1 \Delta_1^{\dagger}\Phi_1 
	+\lambda_{\Phi_1 \Delta3}(\Phi_1^{\dagger}\Phi_1)Tr(\Delta_2^{\dagger}\Delta_2) 	+ \lambda_{\Phi_1 \Delta4} \Phi_1^{\dagger}\Delta_2 \Delta_2^{\dagger}\Phi_1  \\ \nn 
		\end{align}
	\begin{align}
	&	+\lambda_{\Phi_1 \Delta5}(\Phi_1^{\dagger}\Phi_1)Tr(\Delta_3^{\dagger}\Delta_3)
	+ \lambda_{\Phi_1 \Delta6} \Phi_1^{\dagger}\Delta_3 \Delta_3^{\dagger}\Phi_1 
	+\lambda_{\Delta12}(Tr(\Delta_1^{\dagger}\Delta_2))^2   + \lambda_{\Delta21}Tr(\Delta_1^{\dagger}\Delta_2)^2 +\lambda_{\Delta13}(Tr(\Delta_1^{\dagger}\Delta_3))^2\nn \\ &    + \lambda_{\Delta31}Tr(\Delta_1^{\dagger}\Delta_3)^2 
	+\lambda_{\Delta23}(Tr(\Delta_2^{\dagger}\Delta_3))^2
	+ \lambda_{\Delta32}Tr(\Delta_2^{\dagger}\Delta_3)^2 \, , \nn
\end{align}
where, the field definitions for the Higgs doublet and the three triplets are given as follows;
\begin{center}
	$	\Phi_1
	= \left(\begin{array}{c}
		G^+   \\
		\frac{1}{\sqrt{2}}(v_\phi+\rho_1+i G^0)  \end{array}\right) $, \qquad \qquad
	$\Delta_1 =\frac{1}{\sqrt{2}} \left(
	\begin{array}{cc}
		\delta_1^+ & \sqrt{2} \delta_1^{++} \\
		(v_{\Delta1} + \delta_1^0 + i \eta_1^0) & -\delta_1^+ \\
	\end{array}
	\right)$, \\
	\quad 	$\Delta_2 =\frac{1}{\sqrt{2}} \left(
	\begin{array}{cc}
		\delta_2^+ & \sqrt{2} \delta_2^{++} \\
		(v_{\Delta2} + \delta_2^0 + i \eta_2^0) & -\delta_2^+ \\
	\end{array}
	\right)$, \qquad $\Delta_3 =\frac{1}{\sqrt{2}} \left(
	\begin{array}{cc}
		\delta_3^+ & \sqrt{2} \delta_3^{++} \\
		(v_{\Delta3} + \delta_3^0 + i \eta_3^0) & -\delta_3^+ \\
	\end{array}
	\right).$
\end{center}

The $\mu$ term is possible only with the triplet having  zero $U(1)_{L_{\mu}-L_{\tau}}$ charge. Assigning different $U(1)_{L_{\mu}-L_{\tau}}$
charges to the scalar triplets would eliminate several quartic operators from the renormalizable scalar potential due to gauge invariance. This reduction of interaction terms would in turn weaken the renormalization-group running and relax the resulting theoretical constraints. In order to capture the maximal impact of the extended scalar sector on the running behavior and the associated bounds, we therefore adopt the choice that all three triplet scalars carry identical non-zero $U(1)_{L_{\mu}-L_{\tau}}$ charges. With this assignment, the scalar potential contains the full set of allowed renormalizable interactions, providing a conservative and comprehensive framework for our analysis. After writing the scalar potential, we can compute the theoretical bounds on the parameter space of the model from the vacuum stability and the perturbative unitarity in the next section.

\section{Stability Conditions}\label{stability}
The scalar potential discussed in the previous section should be bounded from below in all possible directions, and hence, 
the tree-level stability conditions for the SM Higgs doublet extended with three non-zero hypercharge triplets imposed with $U(1)_{L_{\mu}-L_{\tau}}$ symmetry are given as follows;
\bea
\lambda_{\Phi1}>0\, , k1= \lambda_{\Delta1} + \lambda_{\Delta2} > 0 \, , k2=  \lambda_{\Delta3} + \lambda_{\Delta4} > 0\, , k3= \lambda_{\Delta5} + \lambda_{\Delta6} > 0\, , \nn
\eea
\hspace*{+8.2cm} and
\bea
\lambda_{\Delta1} + \frac{1}{2}\lambda_{\Delta2} >0\, , a1=\sqrt{\lambda_{\Phi1}(\lambda_{\Delta1} + \lambda_{\Delta2})} + \lambda_{\Phi_1 \Delta1} >0\, ,  a2=\sqrt{\lambda_{\Phi1}(\lambda_{\Delta1} + \lambda_{\Delta2})} + \lambda_{\Phi_1 \Delta1} + \lambda_{\Phi_1 \Delta2}>0, \nn \\ 
\lambda_{\Delta3} + \frac{1}{2}\lambda_{\Delta4} >0\, , a3= \sqrt{\lambda_{\Phi1}(\lambda_{\Delta3} + \lambda_{\Delta4})} + \lambda_{\Phi_1 \Delta3} >0\, , a4= \sqrt{\lambda_{\Phi1}(\lambda_{\Delta3} + \lambda_{\Delta4})} + \lambda_{\Phi_1 \Delta3} + \lambda_{\Phi_1 \Delta4}>0, \nn  \\  
\lambda_{\Delta5} + \frac{1}{2}\lambda_{\Delta6} > 0\, , a5= \sqrt{\lambda_{\Phi1}(\lambda_{\Delta5} + \lambda_{\Delta6})} + \lambda_{\Phi_1 \Delta5} >0\, , a6= \sqrt{\lambda_{\Phi1}(\lambda_{\Delta5} + \lambda_{\Delta6})} + \lambda_{\Phi_1 \Delta5} + \lambda_{\Phi_1 \Delta6}>0, \nn   
\eea
\hspace*{+8.2cm} and
\bea
b1= \sqrt{(\lambda_{\Delta1} + \lambda_{\Delta2})(\lambda_{\Delta3} + \lambda_{\Delta4})}+ \lambda_{\Delta12} >0, \qquad b2= \sqrt{(\lambda_{\Delta1} + \lambda_{\Delta2})(\lambda_{\Delta3} + \lambda_{\Delta4})}+ \lambda_{\Delta12} + \lambda_{\Delta21}  >0, \nn \\ b3=\sqrt{(\lambda_{\Delta1} + \lambda_{\Delta2})(\lambda_{\Delta5} + \lambda_{\Delta6})}+ \lambda_{\Delta13} >0, \qquad b4= \sqrt{(\lambda_{\Delta1} + \lambda_{\Delta2})(\lambda_{\Delta5} + \lambda_{\Delta6})}+ \lambda_{\Delta13} + \lambda_{\Delta31} >0,
\nn \\
b5=\sqrt{(\lambda_{\Delta3} + \lambda_{\Delta4})(\lambda_{\Delta5} + \lambda_{\Delta6})}+ \lambda_{\Delta23} >0, \qquad b6= \sqrt{(\lambda_{\Delta3} + \lambda_{\Delta4})(\lambda_{\Delta5} + \lambda_{\Delta6})}+ \lambda_{\Delta23} + \lambda_{\Delta32} >0, \nn 
\eea
\hspace*{+8.2cm} and
\bea
\Big[\lambda_{\Delta2} \sqrt{\lambda_{\Phi1}} \leq |\lambda_{\Phi_1 \Delta2}| \sqrt{\lambda_{\Delta1} + \lambda_{\Delta2}} > 0,  \qquad \rm{or} \qquad 2\lambda_{\Phi_1 \Delta1}+\lambda_{\Phi_1 \Delta2} + \sqrt{(2\lambda_{\Phi1}\lambda_{\Delta2}-\lambda_{\Phi_1 \Delta2}^2)\Big(2\frac{\lambda_{\Delta1}}{\lambda_{\Delta2}}\Big)} > 0\Big], \nn \\
\Big[\lambda_{\Delta4} \sqrt{\lambda_{\Phi1}} \leq |\lambda_{\Phi_1 \Delta4}| \sqrt{\lambda_{\Delta3} + \lambda_{\Delta4}} >0, \qquad \rm{or} \qquad 2\lambda_{\Phi_1 \Delta3}+\lambda_{\Phi_1 \Delta4} + \sqrt{(2\lambda_{\Phi1}\lambda_{\Delta4}-\lambda_{\Phi_1 \Delta4}^2)\Big(2\frac{\lambda_{\Delta3}}{\lambda_{\Delta4}}\Big)} > 0\Big], \nn \\
\Big[\lambda_{\Delta6} \sqrt{\lambda_{\Phi1}} \leq |\lambda_{\Phi_1 \Delta6}| \sqrt{\lambda_{\Delta5} + \lambda_{\Delta6}} >0, \qquad \rm{or} \qquad 2\lambda_{\Phi_1 \Delta5}+\lambda_{\Phi_1 \Delta6} + \sqrt{(2\lambda_{\Phi1}\lambda_{\Delta6}-\lambda_{\Phi_1 \Delta6}^2)\Big(2\frac{\lambda_{\Delta5}}{\lambda_{\Delta6}}\Big)} > 0\Big]. \nn   
\eea
\hspace*{+8.2cm} and
\bea
\Big[\lambda_{\Delta4} \sqrt{\lambda_{\Delta1} + \lambda_{\Delta2}} \leq |\lambda_{\Delta21}| \sqrt{\lambda_{\Delta3} + \lambda_{\Delta4}} > 0,  \quad \rm{or} \quad 2\lambda_{\Delta12}+\lambda_{\Delta21} + \sqrt{(2(\lambda_{\Delta1} + \lambda_{\Delta2})\lambda_{\Delta4}-\lambda_{ \Delta21}^2)\Big(2\frac{\lambda_{\Delta3}}{\lambda_{\Delta4}}\Big)} > 0\Big], \nn \\
\Big[\lambda_{\Delta6} \sqrt{\lambda_{\Delta1} + \lambda_{\Delta2}} \leq |\lambda_{\Delta31}| \sqrt{\lambda_{\Delta5} + \lambda_{\Delta6}} > 0,  \quad \rm{or} \quad 2\lambda_{\Delta13}+\lambda_{\Delta31} + \sqrt{(2(\lambda_{\Delta1} + \lambda_{\Delta2})\lambda_{\Delta6}-\lambda_{ \Delta31}^2)\Big(2\frac{\lambda_{\Delta5}}{\lambda_{\Delta6}}\Big)} > 0\Big], \nn \\
\Big[\lambda_{\Delta6} \sqrt{\lambda_{\Delta3} + \lambda_{\Delta4}} \leq |\lambda_{\Delta32}| \sqrt{\lambda_{\Delta5} + \lambda_{\Delta6}} > 0,  \quad \rm{or} \quad 2\lambda_{\Delta23}+\lambda_{\Delta32} + \sqrt{(2(\lambda_{\Delta3} + \lambda_{\Delta4})\lambda_{\Delta6}-\lambda_{ \Delta31}^2)\Big(2\frac{\lambda_{\Delta5}}{\lambda_{\Delta6}}\Big)} > 0\Big]. \nn   
\eea
\hspace*{+8.2cm} and
\bea
\Big[\sqrt{\lambda_{\Phi1}}(\lambda_{\Delta12}+\lambda_{\Delta13}+\lambda_{\Delta23})+\sqrt{k1}( \lambda_{\Phi_1 \Delta3}+ \lambda_{\Phi_1 \Delta5}+\lambda_{\Delta23} )+ \sqrt{k2}( \lambda_{\Phi_1 \Delta1}+ \lambda_{\Phi_1 \Delta5}+\lambda_{\Delta13} ) \nn \\
+\sqrt{k3}( \lambda_{\Phi_1 \Delta1}+ \lambda_{\Phi_1 \Delta3}+\lambda_{\Delta12} )+\Big[\sqrt{\lambda_{\Phi1}(k1)(k2)(k3)}+\sqrt{(a1 )(a3)(a5)(b1)(b3)(b5)}\Big] >0 \Big], \nn \\
\Big[\sqrt{\lambda_{\Phi1}}(\lambda_{\Delta12}+\lambda_{\Delta21}+\lambda_{\Delta13}+\lambda_{\Delta31}+\lambda_{\Delta23}+\lambda_{\Delta32})+\sqrt{k1}( \lambda_{\Phi_1 \Delta3}+ \lambda_{\Phi_1 \Delta4}+ \lambda_{\Phi_1 \Delta5}+\lambda_{\Phi_1 \Delta6}+\lambda_{\Delta23}+\lambda_{\Delta32} )\nn \\+ \sqrt{k2}( \lambda_{\Phi_1 \Delta1}+ \lambda_{\Phi_1 \Delta2}+ \lambda_{\Phi_1 \Delta5}+\lambda_{\Phi_1 \Delta6}+\lambda_{\Delta13}+\lambda_{\Delta31} ) +
\sqrt{k3}( \lambda_{\Phi_1 \Delta1}+\lambda_{\Phi_1 \Delta2}+ \lambda_{\Phi_1 \Delta3}+\lambda_{\Phi_1 \Delta4}+\lambda_{\Delta12}+\lambda_{\Delta21} )\nn \\
+\Big[\sqrt{\lambda_{\Phi1}(k1)(k2)(k3)}+\sqrt{(a2 )(a4)(a6)(b2)(b4)(b6)}\Big] >0 \Big]. \nn
\eea

The allowed parameter space for the interaction couplings of the triplets with the Higgs from the tree-level stability conditions is given in \autoref{fig:treestab}(a)-(c)-(e). The value of Higgs quartic coupling $\lambda_{\Phi_1}$ is fixed to 0.13 and the values of other quartic couplings are varied from -1.0 to 1.0 for plotting the \autoref{fig:treestab}. The allowed parameter space mostly lies in the positive quadrant of the $[{-1.0,1.0},{-1.0,1.0}]$ plane for all the interaction couplings. The bounded-from-below conditions derived from the scalar potential in \autoref{scalar} consist of several coupled and formally nonlinear inequalities involving the quartic couplings $\lambda_{\Phi_1}, \lambda_{\Delta_i}, \lambda_{\Delta_{ij}}$ and $ \lambda_{\Phi_1 \Delta_i}$, their practical impact on the parameter scan is considerably simpler. In the parameter ranges relevant for \autoref{fig:treestab}, the most restrictive stability requirements typically involve mixed quartic couplings constrained by relations of the form $\lambda_{\Delta_{ij}} \gtrsim - \sqrt{\lambda_{\Delta_i}\lambda_{\Delta_j}}, \lambda_{\Phi_1 \Delta_i} \gtrsim - \sqrt{\lambda_{\Phi_1} \lambda_{\Delta_i}}$. 
	Since the Higgs quartic coupling $\lambda_{\Phi_1}$ is fixed by the measured Higgs mass and the diagonal triplet self-couplings $\lambda_{\Delta_i}$ are restricted to positive and relatively narrow intervals by perturbativity and RGE evolution, the square-root terms vary only weakly over the scan. Consequently, these conditions translate into approximately linear lower bounds on the mixed quartic couplings.
	This leads to the common features observed in all subfigures of \autoref{fig:treestab}, namely a similar lower bound on the horizontal axis and an almost linear behavior of the minimum allowed values on the vertical axis. Such a pattern is therefore not accidental, but reflects the dominance of a small number of leading stability constraints once theoretical consistency conditions are imposed.
	The same reasoning applies to other quartic couplings appearing in the scalar potential. In each two-dimensional projection, only a subset of the full BFB conditions plays a significant role across most of the allowed parameter space, while the remaining inequalities become subleading. As a result, the allowed regions appear simple and nearly linear despite the complicated structure of the complete set of stability conditions. Now, to check the stability of the electroweak vacuum from electroweak scale to the Planck scale, the running of the quartic couplings is computed.The renormalization group equations (RGEs) at one-loop level are given in \autoref{twoloop}. 
	
After discussing the tree-level stability conditions, it is also important to study the validity of the electroweak vacuum stability at higher energy scales. The relevance of higher-order renormalization group effects has been widely discussed in the literature, particularly in studies of near-criticality of the Higgs potential (see, \cite{Degrassi:2012ry}). These works show that while one-loop beta functions capture the leading logarithmic behavior and are sufficient to understand the qualitative trends of the running couplings, they are generally insufficient for quantitatively reliable conclusions when the theory approaches vacuum instability or perturbativity limits.
	In our model, the relevant quartic couplings may evolve close to critical values over a broad range of energy scales. In this regime, even small numerical differences between one-loop and two-loop running can accumulate and lead to noticeable shifts in the scale at which vacuum stability is lost or a Landau pole appears. This behavior is well known from near-criticality analyses of the Standard Model, where higher-order effects play an essential role in determining the instability scale.
	We have verified that the one-loop running reproduces the overall qualitative behavior observed at the two-loop level. However, the precise location of the stability boundaries and the cutoff scale exhibits a non-negligible sensitivity to higher-order corrections. For this reason, while a one-loop analysis is sufficient to illustrate the general features of the RG evolution, the inclusion of two-loop beta functions is necessary for a quantitatively robust assessment of vacuum stability and perturbativity in the present study. Hence, the running of the quartic couplings is taken into account till Planck scale and the allowed parameter space is examined in the below section.

\subsection{Two-loop stability}\label{twoloop}
This section includes one-loop $\beta$-functions for the dimensionless quartic couplings. The detailed expressions for the two-loop $\beta$-functions are given in \autoref{betaf1}. The two-loop $\beta$-functions are computed using package {\tt{SARAH}} \cite{Staub:2013tta}. In the present analysis, the renormalization group evolution is performed assuming a single effective theory, and threshold corrections associated with integrating out the heavy scalar states are not included. This choice was made for simplicity and to allow a transparent comparison between tree-level, one-loop, and two-loop running effects within a unified framework. We agree that, when the invariant masses of the additional scalar fields are of order $\mathcal{O}(1)$ TeV, threshold effects can quantitatively modify the running of the couplings above the mass scale of the new states and may shift the precise position of the Landau pole. However, such effects typically introduce order-one numerical changes rather than altering the qualitative behavior of the running. In particular, the emergence of a Landau pole at an intermediate scale, which limits the validity of the model as an effective theory, remains a robust feature.

\begin{align*}
	\beta_{\lambda_{\Phi_1}} \ =  \ &
	\frac{1}{16\pi^2} \Bigg[+\frac{27}{200} g_{1}^{4} +\frac{9}{20} g_{1}^{2} g_{2}^{2} +\frac{9}{8} g_{2}^{4} -\frac{9}{5} g_{1}^{2} \lambda_{\Phi_1} -9 g_{2}^{2} \lambda_{\Phi_1} +24 \lambda_{\Phi_1}^{2} +3 \lambda_{\Phi_1 \Delta1}^{2} +3 \lambda_{\Phi_1 \Delta1} \lambda_{\Phi_1 \Delta2} +\frac{5}{4} \lambda_{\Phi_1 \Delta2}^{2} \nonumber \\
	& +3 \lambda_{\Phi_1 \Delta3}^{2} +3 \lambda_{\Phi_1 \Delta3} \lambda_{\Phi_1 \Delta4} +\frac{5}{4} \lambda_{\Phi_1 \Delta4}^{2}+3 \lambda_{\Phi_1 \Delta5}^{2} +3 \lambda_{\Phi_1 \Delta5} \lambda_{\Phi_1 \Delta6} +\frac{5}{4} \lambda_{\Phi_1 \Delta6}^{2}+12 \lambda_{\Phi_1} \mbox{Tr}\Big({Y_d  Y_{d}^{\dagger}}\Big)   \nonumber \\ 
	&+4 \lambda_{\Phi_1} \mbox{Tr}\Big({Y_e  Y_{e}^{\dagger}}\Big) +12 \lambda_{\Phi_1} \mbox{Tr}\Big({Y_u  Y_{u}^{\dagger}}\Big) -6 \mbox{Tr}\Big({Y_d  Y_{d}^{\dagger}  Y_d  Y_{d}^{\dagger}}\Big) -2 \mbox{Tr}\Big({Y_e  Y_{e}^{\dagger}  Y_e  Y_{e}^{\dagger}}\Big) -6 \mbox{Tr}\Big({Y_u  Y_{u}^{\dagger}  Y_u  Y_{u}^{\dagger}}\Big) \Bigg]\, , \\
	\beta_{\lambda_{\Phi_1 \Delta2}} \ =  \ &
	\frac{1}{16\pi^2} \Bigg[+\frac{36}{5} g_{1}^{2} g_{2}^{2} -\frac{9}{2} g_{1}^{2} \lambda_{\Phi_1 \Delta2} -\frac{33}{2} g_{2}^{2} \lambda_{\Phi_1 \Delta2} +4 \lambda_{\Phi_1}  \lambda_{\Phi_1 \Delta2} +8 \lambda_{\Phi_1 \Delta1} \lambda_{\Phi_1 \Delta2} +4 \lambda_{\Phi_1 \Delta2}^{2} +4 \lambda_{\Phi_1 \Delta2} \lambda_{\Delta1} \nonumber \\
	&+8 \lambda_{\Phi_1 \Delta2} \lambda_{\Delta2} +2 \lambda_{\Phi_1 \Delta4} \lambda_{\Delta21} +2 \lambda_{\Phi_1 \Delta6} \lambda_{\Delta31}+6 \lambda_{\Phi_1 \Delta2} \mbox{Tr}\Big({Y_d  Y_{d}^{\dagger}}\Big) +2 \lambda_{\Phi_1 \Delta2} \mbox{Tr}\Big({Y_e  Y_{e}^{\dagger}}\Big) +6 \lambda_{\Phi_1 \Delta2} \mbox{Tr}\Big({Y_u  Y_{u}^{\dagger}}\Big) \Bigg]\, , \\
	\beta_{\lambda_{\Phi_1 \Delta1}} \ =  \ &
	\frac{1}{16\pi^2} \Bigg[+\frac{27}{25} g_{1}^{4} -\frac{18}{5} g_{1}^{2} g_{2}^{2} +6 g_{2}^{4} -\frac{9}{2} g_{1}^{2} \lambda_{\Phi_1 \Delta1} -\frac{33}{2} g_{2}^{2} \lambda_{\Phi_1 \Delta1} +12 \lambda_{\Phi_1} \lambda_{\Phi_1 \Delta1} +4 \lambda_{\Phi_1 \Delta1}^{2} +4 \lambda_{\Phi_1} \lambda_{\Phi_1 \Delta2} +\lambda_{\Phi_1 \Delta2}^{2} \nonumber \\
	&+16 \lambda_{\Phi_1 \Delta1} \lambda_{\Delta1} +6 \lambda_{\Phi_1 \Delta2} \lambda_{\Delta1}+12 \lambda_{\Phi_1 \Delta1} \lambda_{\Delta2} +2 \lambda_{\Phi_1 \Delta2} \lambda_{\Delta2} +6 \lambda_{\Phi_1 \Delta3} \lambda_{\Delta12} +3 \lambda_{\Phi_1 \Delta4} \lambda_{\Delta12}+6 \lambda_{\Phi_1 \Delta5} \lambda_{\Delta13}  \nonumber \\ 
	& +3 \lambda_{\Phi_1 \Delta6} \lambda_{\Delta13} +3 \lambda_{\Phi_1 \Delta3} \lambda_{\Delta21} +\frac{1}{2} \lambda_{\Phi_1 \Delta4} \lambda_{\Delta21} +3 \lambda_{\Phi_1 \Delta5} \lambda_{\Delta31} +\frac{1}{2} \lambda_{\Phi_1 \Delta6} \lambda_{\Delta31}+6 \lambda_{\Phi_1 \Delta1} \mbox{Tr}\Big({Y_d  Y_{d}^{\dagger}}\Big) \nonumber \\ 
	& +2 \lambda_{\Phi_1 \Delta1} \mbox{Tr}\Big({Y_e  Y_{e}^{\dagger}}\Big) +6 \lambda_{\Phi_1 \Delta1} \mbox{Tr}\Big({Y_u  Y_{u}^{\dagger}}\Big) \Bigg]\, , \\
	\beta_{\lambda_{\Phi_1 \Delta3}} \ =  \ &
	\frac{1}{16\pi^2} \Bigg[+\frac{27}{25} g_{1}^{4} -\frac{18}{5} g_{1}^{2} g_{2}^{2} +6 g_{2}^{4} -\frac{9}{2} g_{1}^{2} \lambda_{\Phi_1 \Delta3} -\frac{33}{2} g_{2}^{2} \lambda_{\Phi_1 \Delta3} +12 \lambda_{\Phi_1}  \lambda_{\Phi_1 \Delta3} +4 \lambda_{\Phi_1 \Delta3}^{2} +4 \lambda_{\Phi_1}  \lambda_{\Phi_1 \Delta4} +\lambda_{\Phi_1 \Delta4}^{2}\nonumber \\
	&+6 \lambda_{\Phi_1 \Delta1} \lambda_{\Delta12} +3 \lambda_{\Phi_1 \Delta2} \lambda_{\Delta12}+16 \lambda_{\Phi_1 \Delta3} \lambda_{\Delta3} +6 \lambda_{\Phi_1 \Delta4} \lambda_{\Delta3} +3 \lambda_{\Phi_1 \Delta1} \lambda_{\Delta21} +\frac{1}{2} \lambda_{\Phi_1 \Delta2} \lambda_{\Delta21} +12 \lambda_{\Phi_1 \Delta3} \lambda_{\Delta4}  \nonumber \\ 
	&+2 \lambda_{\Phi_1 \Delta4} \lambda_{\Delta4} +6 \lambda_{\Phi_1 \Delta5} \lambda_{\Delta23} +3 \lambda_{\Phi_1 \Delta6} \lambda_{\Delta23} +3 \lambda_{\Phi_1 \Delta5} \lambda_{\Delta32} +\frac{1}{2} \lambda_{\Phi_1 \Delta6} \lambda_{\Delta32} +6 \lambda_{\Phi_1 \Delta3} \mbox{Tr}\Big({Y_d  Y_{d}^{\dagger}}\Big)\nonumber \\ 
	& +2 \lambda_{\Phi_1 \Delta3} \mbox{Tr}\Big({Y_e  Y_{e}^{\dagger}}\Big) +6 \lambda_{\Phi_1 \Delta3} \mbox{Tr}\Big({Y_u  Y_{u}^{\dagger}}\Big) \Bigg]\, , \\
	\beta_{\lambda_{\Phi_1 \Delta5}} \ =  \ &
	\frac{1}{16\pi^2} \Bigg[+\frac{27}{25} g_{1}^{4} -\frac{18}{5} g_{1}^{2} g_{2}^{2} +6 g_{2}^{4} -\frac{9}{2} g_{1}^{2} \lambda_{\Phi_1 \Delta5} -\frac{33}{2} g_{2}^{2} \lambda_{\Phi_1 \Delta5} +12 \lambda_{\Phi_1}  \lambda_{\Phi_1 \Delta5} +4 \lambda_{\Phi_1 \Delta5}^{2} +4 \lambda_{\Phi_1}  \lambda_{\Phi_1 \Delta6} +\lambda_{\Phi_1 \Delta6}^{2}\nonumber \\
	&+6 \lambda_{\Phi_1 \Delta1} \lambda_{\Delta13} +3 \lambda_{\Phi_1 \Delta2} \lambda_{\Delta13} +6 \lambda_{\Phi_1 \Delta3} \lambda_{\Delta23} +3 \lambda_{\Phi_1 \Delta4} \lambda_{\Delta23} +16 \lambda_{\Phi_1 \Delta5} \lambda_{\Delta5} +6 \lambda_{\Phi_1 \Delta6} \lambda_{\Delta5} +3 \lambda_{\Phi_1 \Delta1} \lambda_{\Delta31}\nonumber \\ 
	& +\frac{1}{2} \lambda_{\Phi_1 \Delta2} \lambda_{\Delta31} +3 \lambda_{\Phi_1 \Delta3} \lambda_{\Delta32} +\frac{1}{2} \lambda_{\Phi_1 \Delta4} \lambda_{\Delta32} +12 \lambda_{\Phi_1 \Delta5} \lambda_{\Delta6} +2 \lambda_{\Phi_1 \Delta6} \lambda_{\Delta6}+6 \lambda_{\Phi_1 \Delta5} \mbox{Tr}\Big({Y_d  Y_{d}^{\dagger}}\Big) \nonumber \\ 
	& +2 \lambda_{\Phi_1 \Delta5} \mbox{Tr}\Big({Y_e  Y_{e}^{\dagger}}\Big) +6 \lambda_{\Phi_1 \Delta5} \mbox{Tr}\Big({Y_u  Y_{u}^{\dagger}}\Big) \Bigg]\, , \\
	\beta_{\lambda_{\Phi_1 \Delta4}} \ =  \ &
	\frac{1}{16\pi^2} \Bigg[+\frac{36}{5} g_{1}^{2} g_{2}^{2} -\frac{9}{2} g_{1}^{2} \lambda_{\Phi_1 \Delta4} -\frac{33}{2} g_{2}^{2} \lambda_{\Phi_1 \Delta4} +4 \lambda_{\Phi_1}  \lambda_{\Phi_1 \Delta4} +8 \lambda_{\Phi_1 \Delta3} \lambda_{\Phi_1 \Delta4} +4 \lambda_{\Phi_1 \Delta4}^{2} +4 \lambda_{\Phi_1 \Delta4} \lambda_{\Delta3} \nonumber \\
	&+2 \lambda_{\Phi_1 \Delta2} \lambda_{\Delta21} +8 \lambda_{\Phi_1 \Delta4} \lambda_{\Delta4} +2 \lambda_{\Phi_1 \Delta6} \lambda_{\Delta32} +6 \lambda_{\Phi_1 \Delta4} \mbox{Tr}\Big({Y_d  Y_{d}^{\dagger}}\Big) +2 \lambda_{\Phi_1 \Delta4} \mbox{Tr}\Big({Y_e  Y_{e}^{\dagger}}\Big) +6 \lambda_{\Phi_1 \Delta4} \mbox{Tr}\Big({Y_u  Y_{u}^{\dagger}}\Big) \Bigg]\, , \\
	\beta_{\lambda_{\Phi_1 \Delta6}} \ =  \ &
	\frac{1}{16\pi^2} \Bigg[+\frac{36}{5} g_{1}^{2} g_{2}^{2} -\frac{9}{2} g_{1}^{2} \lambda_{\Phi_1 \Delta6} -\frac{33}{2} g_{2}^{2} \lambda_{\Phi_1 \Delta6} +4  \lambda_{\Phi_1} \lambda_{\Phi_1 \Delta6} +8 \lambda_{\Phi_1 \Delta5} \lambda_{\Phi_1 \Delta6} +4 \lambda_{\Phi_1 \Delta6}^{2} +4 \lambda_{\Phi_1 \Delta6} \lambda_{\Delta5} \nonumber \\
	&+2 \lambda_{\Phi_1 \Delta2} \lambda_{\Delta31} +2 \lambda_{\Phi_1 \Delta4} \lambda_{\Delta32} +8 \lambda_{\Phi_1 \Delta6} \lambda_{\Delta6} +6 \lambda_{\Phi_1 \Delta6} \mbox{Tr}\Big({Y_d  Y_{d}^{\dagger}}\Big) +2 \lambda_{\Phi_1 \Delta6} \mbox{Tr}\Big({Y_e  Y_{e}^{\dagger}}\Big) +6 \lambda_{\Phi_1 \Delta6} \mbox{Tr}\Big({Y_u  Y_{u}^{\dagger}}\Big) \Bigg]\, , \\
	\beta_{\lambda_{\Delta1}} \ =  \ &
	\frac{1}{16\pi^2} \Bigg[+\frac{54}{25} g_{1}^{4} +15 g_{2}^{4} +2 \lambda_{\Phi_1 \Delta1}^{2} +2 \lambda_{\Phi_1 \Delta1} \lambda_{\Phi_1 \Delta2} -24 g_{2}^{2} \lambda_{\Delta1} +28 \lambda_{\Delta1}^{2} -\frac{36}{5} g_{1}^{2} \Big(g_{2}^{2} + \lambda_{\Delta1}\Big)+24 \lambda_{\Delta1} \lambda_{\Delta2} \nonumber \\ 
	& +6 \lambda_{\Delta2}^{2} +3 \lambda_{\Delta12}^{2} +3 \lambda_{\Delta13}^{2}+3 \lambda_{\Delta12} \lambda_{\Delta21} +\frac{1}{4} \lambda_{\Delta21}^{2} +3 \lambda_{\Delta13} \lambda_{\Delta31} +\frac{1}{4} \lambda_{\Delta31}^{2} \Bigg]\, , \\
	\beta_{\lambda_{\Delta2}} \ =  \ &
	\frac{1}{16\pi^2} \Bigg[18 \lambda_{\Delta2}^{2}  -24 g_{2}^{2} \lambda_{\Delta2}  + 24 \lambda_{\Delta1} \lambda_{\Delta2}  -6 g_{2}^{4}  + \frac{36}{5} g_{1}^{2} \Big(2 g_{2}^{2}  - \lambda_{\Delta2} \Big) + \lambda_{\Phi_1 \Delta2}^{2} + \lambda_{\Delta21}^{2} + \lambda_{\Delta31}^{2} \Bigg]\, , \\
	\beta_{\lambda_{\Delta3}} \ =  \ &
	\frac{1}{16\pi^2} \Bigg[+\frac{54}{25} g_{1}^{4} +15 g_{2}^{4} +2 \lambda_{\Phi_1 \Delta3}^{2} +2 \lambda_{\Phi_1 \Delta3} \lambda_{\Phi_1 \Delta4} +3 \lambda_{\Delta12}^{2} -24 g_{2}^{2} \lambda_{\Delta3} +28 \lambda_{\Delta3}^{2} -\frac{36}{5} g_{1}^{2} \Big(g_{2}^{2} + \lambda_{\Delta3}\Big) \nonumber \\ 
\end{align*}
\begin{align*}
	&+3 \lambda_{\Delta12} \lambda_{\Delta21} +\frac{1}{4} \lambda_{\Delta21}^{2} +24 \lambda_{\Delta3} \lambda_{\Delta4}+6 \lambda_{\Delta4}^{2} +3 \lambda_{\Delta23}^{2} +3 \lambda_{\Delta23} \lambda_{\Delta32} +\frac{1}{4} \lambda_{\Delta32}^{2}\Bigg]\, , \\
	\beta_{\lambda_{\Delta4}} \ =  \ &
	\frac{1}{16\pi^2} \Bigg[18 \lambda_{\Delta4}^{2}  -24 g_{2}^{2} \lambda_{\Delta4}  + 24 \lambda_{\Delta3} \lambda_{\Delta4}  -6 g_{2}^{4}  + \frac{36}{5} g_{1}^{2} \Big(2 g_{2}^{2}  - \lambda_{\Delta4} \Big) + \lambda_{\Phi_1 \Delta4}^{2} + \lambda_{\Delta21}^{2} + \lambda_{\Delta32}^{2} \Bigg]\, , \\
	\beta_{\lambda_{\Delta5}} \ =  \ &
	\frac{1}{16\pi^2} \Bigg[+\frac{54}{25} g_{1}^{4} +15 g_{2}^{4} +2 \lambda_{\Phi_1 \Delta5}^{2} +2 \lambda_{\Phi_1 \Delta5} \lambda_{\Phi_1 \Delta6} +3 \lambda_{\Delta13}^{2} +3 \lambda_{\Delta23}^{2} -24 g_{2}^{2} \lambda_{\Delta5} +28 \lambda_{\Delta5}^{2} -\frac{36}{5} g_{1}^{2} \Big(g_{2}^{2} + \lambda_{\Delta5}\Big) \nonumber \\ 
	&+3 \lambda_{\Delta13} \lambda_{\Delta31} +\frac{1}{4} \lambda_{\Delta31}^{2}+3 \lambda_{\Delta23} \lambda_{\Delta32} +\frac{1}{4} \lambda_{\Delta32}^{2} +24 \lambda_{\Delta5} \lambda_{\Delta6} +6 \lambda_{\Delta6}^{2} \Bigg]\, , \\
	\beta_{\lambda_{\Delta6}} \ =  \ &
	\frac{1}{16\pi^2} \Bigg[18 \lambda_{\Delta6}^{2}  -24 g_{2}^{2} \lambda_{\Delta6}  + 24 \lambda_{\Delta5} \lambda_{\Delta6}  -6 g_{2}^{4}  + \frac{36}{5} g_{1}^{2} \Big(2 g_{2}^{2}  - \lambda_{\Delta6} \Big) + \lambda_{\Phi_1 \Delta6}^{2} + \lambda_{\Delta31}^{2} + \lambda_{\Delta32}^{2} \Bigg]\, , \\
	\beta_{\lambda_{\Delta12}} \ =  \ &
	\frac{1}{16\pi^2} \Bigg[+\frac{108}{25} g_{1}^{4} +30 g_{2}^{4} +4 \lambda_{\Phi_1 \Delta1} \lambda_{\Phi_1 \Delta3} +2 \lambda_{\Phi_1 \Delta2} \lambda_{\Phi_1 \Delta3} +2 \lambda_{\Phi_1 \Delta1} \lambda_{\Phi_1 \Delta4} -24 g_{2}^{2} \lambda_{\Delta12} +16 \lambda_{\Delta1} \lambda_{\Delta12}\nonumber \\ 
	& +12 \lambda_{\Delta2} \lambda_{\Delta12} +4 \lambda_{\Delta12}^{2} -\frac{36}{5} g_{1}^{2} \Big(2 g_{2}^{2}  + \lambda_{\Delta12}\Big)+16 \lambda_{\Delta12} \lambda_{\Delta3} +6 \lambda_{\Delta1} \lambda_{\Delta21} +2 \lambda_{\Delta2} \lambda_{\Delta21} +6 \lambda_{\Delta3} \lambda_{\Delta21} \nonumber \\
	&+\frac{3}{2} \lambda_{\Delta21}^{2} +12 \lambda_{\Delta12} \lambda_{\Delta4} +2 \lambda_{\Delta21} \lambda_{\Delta4} +6 \lambda_{\Delta13} \lambda_{\Delta23} +3 \lambda_{\Delta23} \lambda_{\Delta31} +3 \lambda_{\Delta13} \lambda_{\Delta32} +\frac{1}{2} \lambda_{\Delta31} \lambda_{\Delta32} \Bigg]\, , \\
	\beta_{\lambda_{\Delta21}} \ =  \ &
	\frac{1}{16\pi^2} \Bigg[-12 g_{2}^{4} +2 \lambda_{\Phi_1 \Delta2} \lambda_{\Phi_1 \Delta4} +\frac{36}{5} g_{1}^{2} \Big(4 g_{2}^{2}  - \lambda_{\Delta21} \Big)-24 g_{2}^{2} \lambda_{\Delta21} +4 \lambda_{\Delta1} \lambda_{\Delta21} +8 \lambda_{\Delta2} \lambda_{\Delta21} +8 \lambda_{\Delta12} \lambda_{\Delta21}\nonumber \\ 
	& +4 \lambda_{\Delta3} \lambda_{\Delta21} +3 \lambda_{\Delta21}^{2} +8 \lambda_{\Delta21} \lambda_{\Delta4} +2 \lambda_{\Delta31} \lambda_{\Delta32} \Bigg]\, , \\
	\beta_{\lambda_{\Delta13}} \ =  \ &
	\frac{1}{16\pi^2} \Bigg[+\frac{108}{25} g_{1}^{4} +30 g_{2}^{4} +4 \lambda_{\Phi_1 \Delta1} \lambda_{\Phi_1 \Delta5} +2 \lambda_{\Phi_1 \Delta2} \lambda_{\Phi_1 \Delta5} +2 \lambda_{\Phi_1 \Delta1} \lambda_{\Phi_1 \Delta6} -24 g_{2}^{2} \lambda_{\Delta13} +16 \lambda_{\Delta1} \lambda_{\Delta13} \nonumber \\ 
	&+12 \lambda_{\Delta2} \lambda_{\Delta13} +4 \lambda_{\Delta13}^{2} -\frac{36}{5} g_{1}^{2} \Big(2 g_{2}^{2}  + \lambda_{\Delta13}\Big)+6 \lambda_{\Delta12} \lambda_{\Delta23} +3 \lambda_{\Delta21} \lambda_{\Delta23} +16 \lambda_{\Delta13} \lambda_{\Delta5} +6 \lambda_{\Delta1} \lambda_{\Delta31} \nonumber \\
	&+2 \lambda_{\Delta2} \lambda_{\Delta31} +6 \lambda_{\Delta5} \lambda_{\Delta31} +\frac{3}{2} \lambda_{\Delta31}^{2} +3 \lambda_{\Delta12} \lambda_{\Delta32} +\frac{1}{2} \lambda_{\Delta21} \lambda_{\Delta32} +12 \lambda_{\Delta13} \lambda_{\Delta6} +2 \lambda_{\Delta31} \lambda_{\Delta6} \Bigg]\, , \\
	\beta_{\lambda_{\Delta31}} \ =  \ &
	\frac{1}{16\pi^2} \Bigg[-12 g_{2}^{4} +2 \lambda_{\Phi_1 \Delta2} \lambda_{\Phi_1 \Delta6} +\frac{36}{5} g_{1}^{2} \Big(4 g_{2}^{2}  - \lambda_{\Delta31} \Big)-24 g_{2}^{2} \lambda_{\Delta31} +4 \lambda_{\Delta1} \lambda_{\Delta31} +8 \lambda_{\Delta2} \lambda_{\Delta31} +8 \lambda_{\Delta13} \lambda_{\Delta31}  \nonumber \\ 
	&+4 \lambda_{\Delta5} \lambda_{\Delta31} +3 \lambda_{\Delta31}^{2} +2 \lambda_{\Delta21} \lambda_{\Delta32}+8 \lambda_{\Delta31} \lambda_{\Delta6} \Bigg]\, , \\
	\beta_{\lambda_{\Delta23}} \ =  \ &
	\frac{1}{16\pi^2} \Bigg[+\frac{108}{25} g_{1}^{4} +30 g_{2}^{4} +4 \lambda_{\Phi_1 \Delta3} \lambda_{\Phi_1 \Delta5} +2 \lambda_{\Phi_1 \Delta4} \lambda_{\Phi_1 \Delta5} +2 \lambda_{\Phi_1 \Delta3} \lambda_{\Phi_1 \Delta6} +6 \lambda_{\Delta12} \lambda_{\Delta13} +3 \lambda_{\Delta13} \lambda_{\Delta21}  \nonumber \\ 
	&-24 g_{2}^{2} \lambda_{\Delta23} +16 \lambda_{\Delta3} \lambda_{\Delta23} +12 \lambda_{\Delta4} \lambda_{\Delta23} +4 \lambda_{\Delta23}^{2}-\frac{36}{5} g_{1}^{2} \Big(2 g_{2}^{2}  + \lambda_{\Delta23}\Big)+16 \lambda_{\Delta23} \lambda_{\Delta5} +3 \lambda_{\Delta12} \lambda_{\Delta31} \nonumber \\
	& +\frac{1}{2} \lambda_{\Delta21} \lambda_{\Delta31} +6 \lambda_{\Delta3} \lambda_{\Delta32} +2 \lambda_{\Delta4} \lambda_{\Delta32} +6 \lambda_{\Delta5} \lambda_{\Delta32} +\frac{3}{2} \lambda_{\Delta32}^{2} +12 \lambda_{\Delta23} \lambda_{\Delta6} +2 \lambda_{\Delta32} \lambda_{\Delta6} \Bigg]\, , \\
	\beta_{\lambda_{\Delta32}} \ =  \ &
	\frac{1}{16\pi^2} \Bigg[-12 g_{2}^{4} +2 \lambda_{\Phi_1 \Delta4} \lambda_{\Phi_1 \Delta6} +2 \lambda_{\Delta21} \lambda_{\Delta31} +\frac{36}{5} g_{1}^{2} \Big(4 g_{2}^{2}  - \lambda_{\Delta32} \Big)-24 g_{2}^{2} \lambda_{\Delta32} +4 \lambda_{\Delta3} \lambda_{\Delta32} +8 \lambda_{\Delta4} \lambda_{\Delta32} \nonumber \\ 
	& +8 \lambda_{\Delta23} \lambda_{\Delta32} +4 \lambda_{\Delta5} \lambda_{\Delta32} +3 \lambda_{\Delta32}^{2}+8 \lambda_{\Delta32} \lambda_{\Delta6} \Bigg]\, . \\
\end{align*}

\begin{figure}[H]
	\begin{center}
		\mbox{\subfigure[ ]{\includegraphics[width=0.5\linewidth,angle=-0]{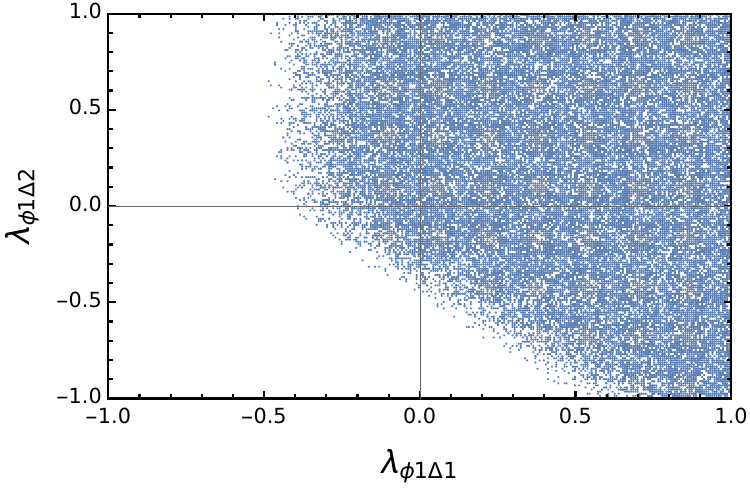}\label{f1a}}
			\subfigure[]{\includegraphics[width=0.5\linewidth,angle=-0]{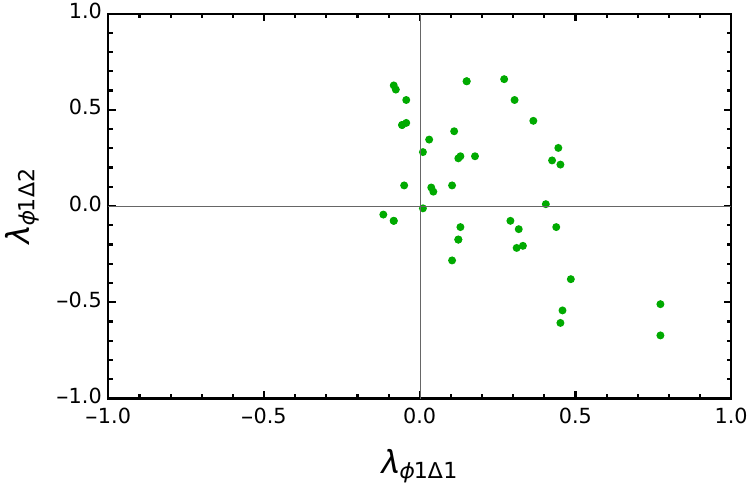}\label{f2a}}}
		\mbox{\subfigure[ ]{\includegraphics[width=0.5\linewidth,angle=-0]{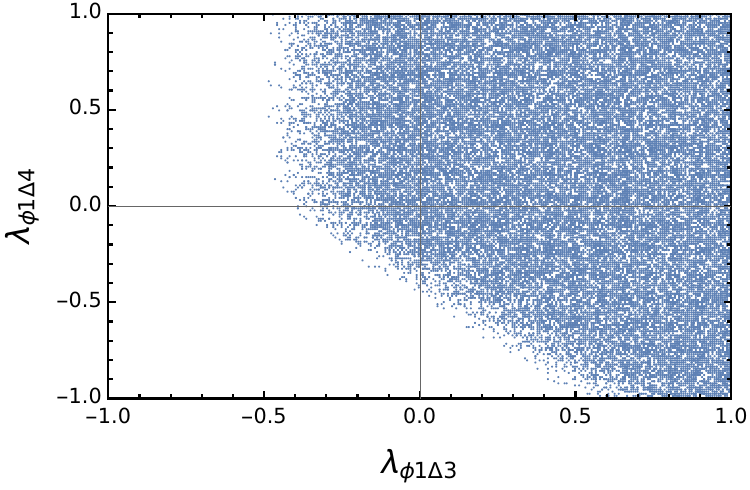}\label{f3a}}
			\subfigure[]{\includegraphics[width=0.5\linewidth,angle=-0]{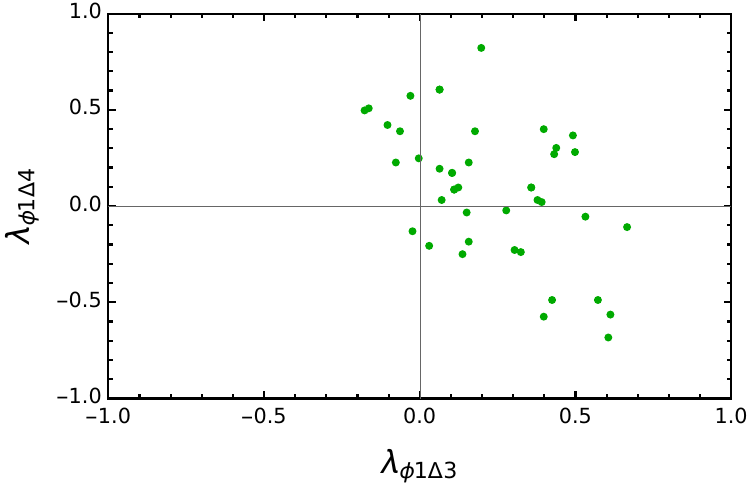}\label{f4a}}}
		\mbox{\subfigure[ ]{\includegraphics[width=0.5\linewidth,angle=-0]{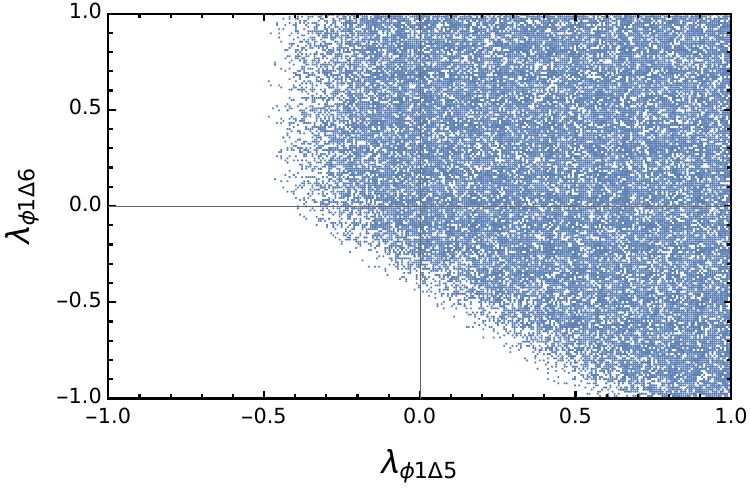}\label{f5a}}
			\subfigure[]{\includegraphics[width=0.5\linewidth,angle=-0]{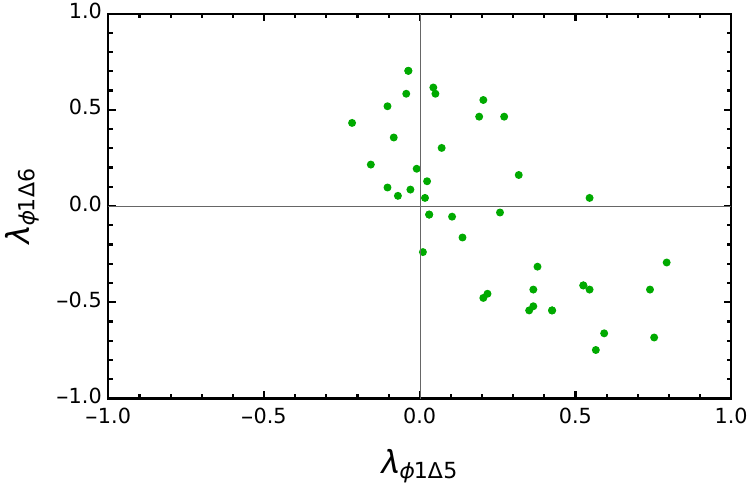}\label{f6a}}}
		\caption{Parameter space satisfying the tree-level stability conditions given in \autoref{stability} in \autoref{fig:treestab}(a)-(c)-(e) and same stability conditions using running couplings with two-loop $\beta$-functions in \autoref{fig:treestab}(b)-(d)-(f) .}\label{fig:treestab}
	\end{center}
\end{figure}

The values of the quartic couplings are fixed at the electroweak scale, and the running of quartic couplings with the energy scale is considered using two-loop $\beta$-functions satisfying the stability conditions as discussed previously. It is expected that the addition of triplet degrees of freedom contributes positively, and enhances the stability of the electroweak vacuum. But in contrary, the positive degrees of freedom makes the quartic couplings hit the Landau pole at much lower energy scales. Here, the interaction quartic couplings are restricted to lower values as depicted in \autoref{fig:treestab}(b)-(d)-(f). Further higher values of the quartic couplings leads to the appearance of Landau pole below Planck scale before the instability. Hence, the allowed parameter space for the interaction quartic couplings is reduced considering the stability till Planck scale. The perturbative unitarity constraints where any of the coupling diverges at a particular energy scale are given as;
\begin{align}
	\left|\lambda_i\right|  \leq 4 \pi \qquad
	\left|g_j\right|  \leq 4 \pi \qquad \left|y_k\right|  \leq 4\pi.
\end{align}

Here, the large positive contribution from the quartic couplings makes it impossible to achieve the perturbative unitarity till Planck scale. The perturbative unitarity is achieved only till $10^{12}$ GeV along with the stability constraints. In our analysis, the perturbative validity of the model is assessed through the renormalization group evolution of all gauge, Yukawa, and scalar quartic couplings. We define the perturbativity cutoff as the energy scale at which at least one of the running couplings exceeds the perturbative regime.
	For the benchmark points considered in this work, we find that all couplings remain perturbative up to scales of order $\mu \sim 10^{12}$ GeV. Beyond this scale, one or more scalar quartic couplings grow rapidly, signaling the breakdown of perturbativity and the onset of strong coupling behavior. We therefore identify $10^{12}$ GeV as the effective upper limit of validity of the model for the benchmark scenarios studied in the later sections.
	Since the dynamics of the electroweak phase transition occurs at temperatures around the electroweak scale, which is many orders of magnitude below this cutoff, our results for the phase transition and the associated phenomenology are not affected by the loss of perturbativity at higher energies.

The white regions in \autoref{fig:treestab}(b)-(d)-(f) correspond to parameter points that are genuinely excluded by the theoretical constraints imposed in our analysis, rather than being artifacts of an insufficient sampling resolution. The scan was performed by uniformly generating points within the specified ranges of the quartic couplings and subsequently applying the bounded-from-below, perturbativity, and RGE-running constraints. Points failing any of these conditions were discarded, which leads to empty regions in the projected parameter space.
We have explicitly checked that increasing the density of scanned points does not qualitatively change the shape of the allowed regions shown in \autoref{fig:treestab}(b)-(d)-(f). The apparent sparsity of colored points is therefore a visualization issue arising from projecting a high-dimensional parameter space onto two dimensions, rather than an indication of incomplete coverage.

After achieving the allowed parameter space from Planck scale stability and perturbative unitarity, it is interesting to check the allowed parameter space for the first-order electroweak phase transition in the next section.

\section{Electroweak phase transition}\label{ewpt}
As the temperature drops, a transition takes place, leading to the broken phase at zero temperature. This transition starts with the early Universe at high temperature, or the symmetric phase. The process of moving from the symmetric phase to the broken phase involves the nucleation of bubbles. Specifically, below the critical temperature $T_c$, the bubbles of the broken phase begin to form in the sea of symmetric phase and continue to do so until the complete symmetric phase transitions into the broken phase. It is possible for this phase transition to be first- or second-order. A barrier between the symmetric phase and the broken phase, which is produced by the cubic term in the potential, is necessary for the strongly first-order phase transition. The tree-level scalar potential in \autoref{scalar} receives additional one-loop contribution at zero temperature from Coleman Weinberg potential and one-loop contribution from finite-temperature potential including thermal-loop using resummation as given below \cite{PhysRevD.7.1888};
\bea\label{eq:4.1}
V_{1-loop}^{\rm CW} & = & \frac{1}{(64\pi)^2}\sum_{i=B,F}(-1)^{F_i} n_i \hat{m_i}^4\Big[\log\Big(\frac{\widetilde{m}_i^2}{\mu^2}\Big)-k_i\Big]. \\
V_{\rm 1-loop}^{T\neq 0} & = & \sum_{\substack{i=W,G^0,Z,h,H1^{\pm\pm}, H2^{\pm\pm},H3^{\pm\pm}, \\   H1^{\pm}, H2^{\pm},H3^{\pm}, A1,H1, A2,H2, A3,H3 }} \frac{n_i T^4}{ 2\pi^2}J_B\Big(\frac{\hat{m}_i^2}{T^2}\Big) + \frac{n_tT^4}{2 \pi^2} J_F\Big(\frac{\hat{m}_t^2}{T^2}\Big), \label{eq:4.2}\\ 
V_{\rm ring}^{T\neq 0} & = & \sum_{\substack{i=W,G^0,Z,h,H1^{\pm\pm}, H2^{\pm\pm},H3^{\pm\pm}, \\   H1^{\pm}, H2^{\pm},H3^{\pm}, A1,H1, A2,H2, A3,H3 }} \frac{n_iT^4}{12 \pi}\Big\{\Big[\frac{\hat{m}_i^2}{T^2}\Big]^{3/2}-\Big[\frac{\widetilde{m}_i^2 }{T^2}\Big]^{3/2}\Big\}\label{eq:4.3},
\eea
where, $\hat{m}_i^2$ and $\widetilde{m}_i^2$ are the field-dependent masses and thermally corrected masses, which include contributions from the Daisy corrections resuming hard thermal loops, respectively. The thermally corrected masses are defined as given below;
\bea
\widetilde{m}_i^2 = \widetilde{m}_i^2(h_1; T) = \hat{m}_i^2(h_1) + \Pi_i T^2,
\eea
where $\Pi_i's$ are the Daisy coefficients, which are zero only for the fermionic fields and only get contribution from the bosonic fields. Since the triplet vev is very small as constrained from the $\rho$ parameter, it is enough to consider the phase transition only in the Higgs direction. In that case the mass expressions for the Higgs and the triplet degrees of freedom are given as follows;

\bea
m^2_{H1^{\pm\pm}}  & = & m_{\Delta1}^2 + \frac{1}{2}\lambda_{\Phi_1 \Delta1} v_{\phi}^2, \nn \\
m^2_{H2^{\pm\pm}}  & = & m_{\Delta2}^2 + \frac{1}{2}\lambda_{\Phi_1 \Delta3} v_{\phi}^2, \nn \\ \nn
\eea
\bea\label{eq:4.5}
m^2_{H3^{\pm\pm}}  & = & m_{\Delta3}^2 + \frac{1}{2}\lambda_{\Phi_1 \Delta5} v_{\phi}^2, \nn \\
m^2_{H1^{\pm}}  & = & m_{\Delta1}^2 + \frac{1}{2}(\lambda_{\Phi_1 \Delta1}+\frac{1}{2}\lambda_{\Phi_1 \Delta2}) v_{\phi}^2, \nn \\
m^2_{H2^{\pm}}  & = & m_{\Delta2}^2 + \frac{1}{2}(\lambda_{\Phi_1 \Delta3}+\frac{1}{2}\lambda_{\Phi_1 \Delta4}) v_{\phi}^2, \nn \\ \nn
m^2_{H3^{\pm}}  & = & m_{\Delta3}^2 + \frac{1}{2}(\lambda_{\Phi_1 \Delta5}+\frac{1}{2}\lambda_{\Phi_1 \Delta6}) v_{\phi}^2, \nn \\ \nn
m^2_{A1/H1}  & = & m_{\Delta1}^2 + \frac{1}{2}(\lambda_{\Phi_1 \Delta1}+\lambda_{\Phi_1 \Delta2}) v_{\phi}^2, \nn 
\\
m^2_{A2/H2}  & = & m_{\Delta2}^2 + \frac{1}{2}(\lambda_{\Phi_1 \Delta3}+\lambda_{\Phi_1 \Delta4}) v_{\phi}^2, \nn 
\\
m^2_{A3/H3}  & = & m_{\Delta3}^2 + \frac{1}{2}(\lambda_{\Phi_1 \Delta5}+\lambda_{\Phi_1 \Delta6}) v_{\phi}^2. 
\eea

Rewriting the scalar potential in \autoref{scalar} in terms of background fields $h_1$, $h_2$, $h_3$ and $h_4$ for the Higgs doublet and three triplets at high temperature, the mass expressions in \autoref{eq:4.5} are  written in terms of the background Higgs field as follows;

\bea
\hat{m}_{h}^2 = 3\lambda_{\Phi1} h_1^2 - m_{\Phi_1}^2, \qquad\hat{m}_{G^0}^2 = \lambda_{\Phi1} h_1^2 - m_{\Phi_1}^2, \qquad \hat{m}_{W}^2 = \frac{g_2^2}{4} h_1^2, \qquad \hat{m}_{Z}^2 = \frac{g_2^2 + g_1^2}{4}h_1^2, \qquad \hat{m}_t^2 =\frac{y_t^2}{2}h_1^2, \nn \\ 
m^2_{H1^{\pm\pm}}   =  m_{\Delta1}^2 + \frac{1}{2}\lambda_{\Phi_1 \Delta1} h_{1}^2, \qquad m^2_{H2^{\pm\pm}}   =  m_{\Delta2}^2 + \frac{1}{2}\lambda_{\Phi_1 \Delta3} h_{1}^2, \qquad m^2_{H3^{\pm\pm}}   =  m_{\Delta3}^2 + \frac{1}{2}\lambda_{\Phi_1 \Delta5} h_{1}^2, \nn \\
m^2_{H1^{\pm}}   =  m_{\Delta1}^2 + \frac{1}{2}(\lambda_{\Phi_1 \Delta1}+\frac{1}{2}\lambda_{\Phi_1 \Delta2}) h_{1}^2,  \qquad m^2_{H2^{\pm}}   =  m_{\Delta2}^2 + \frac{1}{2}(\lambda_{\Phi_1 \Delta3}+\frac{1}{2}\lambda_{\Phi_1 \Delta4}) h_{1}^2, \nn \\ \qquad m^2_{H3^{\pm}}   =  m_{\Delta3}^2 + \frac{1}{2}(\lambda_{\Phi_1 \Delta5}+\frac{1}{2}\lambda_{\Phi_1 \Delta6}) h_{1}^2, \qquad m^2_{A1/H1}   =  m_{\Delta1}^2 + \frac{1}{2}(\lambda_{\Phi_1 \Delta1}+\lambda_{\Phi_1 \Delta2}) h_{1}^2, \nn \\
m^2_{A2/H2}   =  m_{\Delta2}^2 + \frac{1}{2}(\lambda_{\Phi_1 \Delta3}+\lambda_{\Phi_1 \Delta4}) h_{1}^2, \qquad m^2_{A3/H3}   =  m_{\Delta3}^2 + \frac{1}{2}(\lambda_{\Phi_1 \Delta5}+\lambda_{\Phi_1 \Delta6}) h_{1}^2. 
\eea

The masses belonging to the same multiplet would have same thermal corrections, and the corresponding expressions for the Debye coefficients are given below;
\bea\label{eq:4.7}
\Pi_{h} & = & \Big(\frac{g_1^2+3g_2^2}{16} + \frac{\lambda_{\Phi_1}}{2}+ \frac{y_t^2}{4}+\frac{(\lambda_{\Phi_1 \Delta1}+\lambda_{\Phi_1 \Delta3}+\lambda_{\Phi_1 \Delta5)}}{6}+ \frac{(\lambda_{\Phi_1 \Delta2}+\lambda_{\Phi_1 \Delta4}+\lambda_{\Phi_1 \Delta6)}}{8}  \Big)T^2, \nn \\
\Pi_{G^0} & = & \Big(\frac{g_1^2+3g_2^2}{16} + \frac{\lambda_{\Phi_1}}{2}+ \frac{y_t^2}{4}+\frac{(\lambda_{\Phi_1 \Delta1}+\lambda_{\Phi_1 \Delta3}+\lambda_{\Phi_1 \Delta5)}}{6}+ \frac{(\lambda_{\Phi_1 \Delta2}+\lambda_{\Phi_1 \Delta4}+\lambda_{\Phi_1 \Delta6)}}{8}  \Big)T^2, \nn \\
\Pi_{\Delta_1} & = & \Big(  \frac{2\lambda_{\Phi_1 \Delta1}+\lambda_{\Phi_1 \Delta2}}{12}+ \frac{(6\lambda_{\Delta1}+5\lambda_{\Delta2})}{12} +\frac{(\lambda_{\Delta12}+\lambda_{\Delta13})}{6}+\frac{(\lambda_{\Delta21}+\lambda_{\Delta31})}{8}  \Big)T^2, \nn \\
\Pi_{\Delta_2} & = & \Big(  \frac{2\lambda_{\Phi_1 \Delta3}+\lambda_{\Phi_1 \Delta4}}{12}+ \frac{(6\lambda_{\Delta3}+5\lambda_{\Delta4})}{12} +\frac{(\lambda_{\Delta12}+\lambda_{\Delta23})}{6}+\frac{(\lambda_{\Delta21}+\lambda_{\Delta32})}{8}  \Big)T^2, \nn \\
\Pi_{\Delta_3} & = & \Big(  \frac{2\lambda_{\Phi_1 \Delta5}+\lambda_{\Phi_1 \Delta6}}{12}+ \frac{(6\lambda_{\Delta5}+5\lambda_{\Delta6})}{12} +\frac{(\lambda_{\Delta13}+\lambda_{\Delta23})}{6}+\frac{(\lambda_{\Delta31}+\lambda_{\Delta32})}{8}  \Big)T^2, \nn \\
\Pi_{W_L} & = & \frac{31}{12}g_2^2 T^2, \nn 
\eea
\bea
\Pi_{W_T} & = & \Pi_{Z_T}=\Pi_{\gamma_T} =0, \nn \\
\widetilde{m}_{Z_L}^2 & = & \frac{1}{2}\hat{m}_Z^2 +\frac{31}{24}(g_1^2+g_2^2)T^2 + \delta, \nn \\
\widetilde{m}_{\gamma_L}^2 & = & \frac{1}{2}\hat{m}_Z^2 +\frac{31}{24}(g_1^2+g_2^2)T^2 - \delta, \nn 
\eea
where all the particles belonging to the same multiplet have same thermal corrections, except the transverse component of the $W,Z$ and $\gamma$ which receives zero thermal corrections \cite{Comelli:1996vm}. $\widetilde{m}_{Z_L}^2$ and $\widetilde{m}_{\gamma_L}^2$ are the Debye masses for the longitudinal component of $Z$ boson and the photon, and $\delta$ is computed as follows;
\bea
\delta^2= \Big(\frac{1}{2}\hat{m}_z^2+\frac{31}{24}(g_1^2+g_2^2)T^2\Big)^2 - \frac{31}{24}g_1^2 g_2^2 T^2( h_1^2 + \frac{31}{8} T^2).
\eea

After setting up the full temperature corrected one-loop potential, we choose the allowed parameter space from the Planck scale stability  for the phase transition analysis as given in \autoref{tab:int1}- \autoref{tab:int2}. 

\begin{table}[H]
	\centering
	\begin{tabular}{|c|c|c|c|c|c|c|c|c|c|c|}
		\hline
		&$\lambda_{\Phi_1}$&$\lambda_{\Phi1 \Delta1}$&  $\lambda_{\Phi1 \Delta2}$&$\lambda_{\Phi1 \Delta3}$&$\lambda_{\Phi1 \Delta4}$ & $\lambda_{\Phi1 \Delta5}$ & $\lambda_{\Phi1 \Delta6}$ & $\lambda_{\Delta12}$ & $\lambda_{\Delta21}$& $\lambda_{\Delta13}$ 
		\\
		\hline
		BP1 &0.1264&-0.103&0.185&0.239&0.475 & 0.360 & 0.284 & 0.157 & 0.476 & 0.326   \\
		\hline
		BP2 &0.1264&0.287&0.193&-0.090&0.175 & 0.040 & -0.10009 & 0.049 & -0.229 & -0.042   \\
		\hline
		BP3 &0.1264&-0.073&-0.1365&-0.016&0.1817 & 0.376266 & 0.3943 & 0.1122 & -0.1691 & 0.15447  \\
		\hline
		BP4 &0.1264&0.229&-0.101&0.0858&0.332 & 0.350 & -0.0743& 0.460& -0.0949 & -0.1726  \\
		\hline
		BP5 &0.1264&0.0862&0.0671&0.341&-0.4251 & 0.4514 & -0.3888& 0.2859& -0.109 & -0.0848  \\
		\hline
	\end{tabular}
	\caption{Benchmark points chosen from the allowed values satisfying the stability till Planck scale for electroweak phase transition.}
	\label{tab:int1}
\end{table}

\begin{table}[h]
	\centering
	\begin{tabular}{|c|c|c|c|c|c|c|c|c|c|}
		\hline
		&$\lambda_{\Delta31}$&$\lambda_{\Delta23}$&$\lambda_{\Delta32}$& $\lambda_{ \Delta1}$&$\lambda_{ \Delta2}$&$\lambda_{ \Delta3}$ & $\lambda_{ \Delta4}$ & $\lambda_{ \Delta5}$ & $\lambda_{ \Delta6}$ 
		\\
		\hline
		BP1 & -0.473&0.309 &-0.018&0.220&0.032&0.104&-0.0096& -0.0396 & 0.322   \\
		\hline
		BP2 & -0.148  & 0.0416  &0.192&-0.0190&0.150&0.075&0.382 & 0.329 & 0.067 \\
		\hline
		BP3 & -0.3776 & 0.0001&0.4943&0.4985&-0.015734&0.334779&-0.15961 & 0.283303 & -0.108596  \\
		\hline
		BP4 & 0.13708 & -0.02622&-0.1855&-0.0214&0.08057&-0.15077&0.3463 & 0.2467 & 0.310866  \\
		\hline
		BP5 & 0.3742 & -0.1985&0.0967&0.2108&0.3512&0.3714&0.0799 & -0.0922 & 0.4764  \\
		\hline
	\end{tabular}
	\caption{Benchmark points chosen from the allowed values satisfying the stability till Planck scale for electroweak phase transition.}
	\label{tab:int2}
\end{table}

The strength of phase transition $\phi_+(T_c)/T_c$ \cite{Cohen:1993nk, Rubakov:1996vz} for the chosen benchmark points is given in \autoref{tab:int3} for different values of bare mass parameters, where $\phi_+(T_c) = \sqrt{(h_1^l-h_1^h)^2 + (h_2^l - h_2^h)^2 + (h_3^l- h_3^h)^2 + (h_4^l -h_4^h)^2}$ at the critical temperature $T_c$ with the superscript $h$ and $l$ denotes the high-vev and the low-vev for the SM Higgs and three triplet fields. $T_c$ is the critical temperature where two different minima of the potential are degenerate, i.e., $V(0;T_c)= V(\phi_+; T_c)$. The contribution from the SM Higgs doublet and the triplet degrees of freedom to the cubic term is large enough making the transition to be strongly first-order. The strength of phase transition for the variation of bare mass parameters from $50.0$ GeV - $1000.0$ GeV is always strongly first-order varying from $1.51-1.32$ ,$1.43-1.33$, $1.47-1.336$, $1.44-1.313$ and $1.44-1.329$ for BP1, BP2, BP3, BP4 and BP5, respectively. The strength of phase transition reduces with the increase in the mass but it remains strongly first-order till the masses are heavy enough to decouple from the thermal bath and the phase transition strength saturates.

\begin{table}[h]
	\centering
	\begin{tabular}{|c|c|c|c|c|c|}
		\hline
		&BP1&BP2&BP3 & BP4 & BP5
		\\
		\hline
		$m_{\Delta1}/m_{\Delta2}/m_{\Delta3}$(GeV)&$\phi_+(T_c)/T_c$&$\phi_+(T_c)/T_c$& $\phi_+(T_c)/T_c$ & $\phi_+(T_c)/T_c$&$\phi_+(T_c)/T_c$
		\\
		\hline
		50.0 &1.51 &1.43&1.47& 1.44& 1.44\\
		\hline
		100.0 & 1.45  &1.41&1.43& 1.41 & 1.41\\
		\hline
		150.0 & 1.41&1.39&1.41 & 1.39& 1.40\\
		\hline
		200.0 & 1.398&1.38&1.395 & 1.385& 1.389\\
		\hline
		250.0 & 1.387&1.379&1.389 & 1.381& 1.384\\
		\hline
		500.0 & 1.35&1.379&1.385 & 1.369& 1.379\\
		\hline
		1000.0 & 1.32&1.33&1.336 & 1.313& 1.329\\
		\hline
	\end{tabular}
	\caption{Strength of electroweak phase transition for different bare mass parameters in GeV for each of the benchmark points.}
	\label{tab:int3}
\end{table}


This phase transition occurring via bubble nucleation also leads to collision of bubbles which results in shock waves similar to the Gravitational wave signatures (GWs). The intensity of the Gravitational waves (GW) consists of three different contributions; 1) Bubble wall collision \cite{Kosowsky:1991ua, Kosowsky:1992vn, Huber:2008hg, PhysRevLett.69.2026, PhysRevD.49.2837, PhysRevD.77.124015}, 2) Sound waves in the plasma \cite{PhysRevLett.112.041301, Leitao:2012tx, Giblin:2013kea, Giblin:2014qia, PhysRevD.92.123009}, 3) Magnetohydrodynamic turbulence in plasma \cite{PhysRevD.74.063521, PhysRevD.78.043003, Kahniashvili:2008pe, Kahniashvili:2009mf, Caprini:2009yp}, the net contribution combining these three linearly is as follows \cite{Caprini:2015zlo};
\bea
h^2 \Omega_{GW}\simeq h^2 \Omega_{\phi}+h^2 \Omega_{sw}+h^2 \Omega_{turb}.
\eea

The first term is computed using the envelope approximation \cite{Kosowsky:1992vn, Huber:2008hg, PhysRevLett.69.2026} via numerical simulations and is given as;
\bea
h^2 \Omega_{env}(f) = 1.67 \times 10^{-5} \Big(\frac{\beta}{H}\Big)^{-2} \Big(\frac{\kappa_{\phi} \alpha}{1 + \alpha}\Big)^2 \Big(\frac{100}{g_*}\Big)^{1/3} \Big(\frac{0.11 v_w^3}{0.42 + v_w^2}\Big)\frac{3.8 (f/f_{env})^{2.8}}{1+2.8(f/f_{env})^{3.8}},
\eea
with 
\bea
\beta=\Big[HT\frac{d}{dT}\Big(\frac{S_3}{T}\Big)\Big]\Big|_{T_n}.
\eea
$T_n$ is the nucleation temperature at which bubble nucleation begins, $H$ is the Hubble parameter, and $\beta$ indicates the duration of the phase transition.
For the critical bubble in the spherical polar coordinates, the Euclidean action of the background field is defined as $S_3$, and is expressed as follows \cite{Linde:1981zj};
\bea
S_3= 4 \pi \int dr r^2 \Big[\frac{1}{2}(\partial_r \vec{\phi})^2+V_1(T)\Big]. 
\eea
$\alpha, \kappa_\phi, \kappa_v$, and $v_w$ are additional crucial parameters for calculating the GW background.
$\alpha$ denotes the ratio of the vacuum energy density to the radiation bath that is released during the phase transition, which is described as;
\bea
\alpha= \frac{\rho_{vac}}{\rho_{rad}^*},
\eea
where $\rho^*_{rad}=g_*\pi^2T_*^4/30$, $g_*$ is defined as the number of relativistic degrees of freedom in plasma at temperature $T_*$ with $T_*=T_n$ in the absence of reheating.
Other parameters needed for computing the GW frequencies are found as~\cite{Caprini:2015zlo, Shajiee:2018jdq, Kamionkowski:1993fg, Chao:2017vrq, Dev:2019njv, Steinhardt:1981ct};
\bea\label{alp}
\kappa_v & = & \frac{\rho_v}{\rho_{vac}}, \, \qquad  \kappa_\phi=\frac{\rho_\phi}{\rho_{vac}}= 1-\frac{\alpha_\infty}{\alpha}, \, \qquad v_w=\frac{1/\sqrt{3}+\sqrt{\alpha^2+2\alpha/3}}{1+\alpha}, \nn \\
\alpha_{\infty} & = & \frac{30}{24 \pi^2 g_*}\Big(\frac{v_n}{T_n}\Big)^2\Big[6\Big(\frac{m_W}{v}\Big)^2+3\Big(\frac{m_Z}{v}\Big)^2+6\Big(\frac{m_t}{v}\Big)^2\Big].
\eea 
The proportion of vacuum energy that is transformed into the bulk motion of the fluid is defined by $\kappa_v$, while the fraction of vacuum energy that is transformed into the gradient energy of the Higgs-like field is defined by $\kappa_\phi$.
The specified bubble wall velocity of the fluid is $v_w$; the Higgs field velocities at zero temperature and the nucleation temperature are $v, v_n$. The masses of the top quark, $Z$ boson, $W$ boson, and $T_n$ are $m_W, m_Z$, and $m_t$, respectively.
The peak frequency $f_{env}$, which contributes to the GW intensity derived from the bubble collisions, can finally be expressed as follows:
\bea
f_{\rm env}=16.5 \times 10^{-6} Hz \Big(\frac{0.62}{v^2_w-0.1 v_w+1.8}\Big)\Big(\frac{\beta}{H}\Big)\Big(\frac{T_n}{100 \rm GeV}\Big)\Big(\frac{g_*}{100}\Big)^{\frac{1}{6}}.
\eea

Second, the following represents the contribution of sound waves in the plasma to the GW intensity:
\bea
h^2 \Omega_{\rm SW}=2.65\times 10^{-6}\Big(\frac{\beta}{H}\Big)^{-1}v_w \Big(\frac{\kappa_v \alpha}{1+\alpha}\Big)^2 \Big(\frac{g_*}{100}\Big)^{-\frac{1}{3}}\Big(\frac{f}{f_{\rm SW}}\Big)^3
\left[\frac{7}{4+3\Big(\frac{f}{f_{\rm SW}}\Big)^2}\right]^2,
\eea
where the fraction of latent heat that is transferred to the bulk motion of the fluid, denoted by the parameter $\kappa_v$, which was previously given in \autoref{alp}, may now be rewritten as;
\bea \label{eq:4.22}
\kappa_v=\frac{\alpha_{\infty}}{\alpha}\Big[\frac{\alpha_{\infty}}{0.73+0.083 \sqrt{\alpha_{\infty}}+\alpha_{\infty}}\Big].
\eea
The peak frequency contribution computed from the sound wave mechanisms, $f_{\rm SW}$ to the GW spectrum is as follows;
\bea
f_{\rm SW}=1.9\times 10^{-5}Hz \Big(\frac{1}{v_w}\Big)\Big(\frac{\beta}{H}\Big)\Big(\frac{T_n}{100 \rm GeV}\Big)\Big(\frac{g_*}{100}\Big)^{\frac{1}{6}}.
\eea
Finally, \cite{Hogan:1986qda} provides the contribution to the GW spectrum from the Magnetohydrodynamic turbulence ;
\bea
h^2 \Omega_{\rm turb}=3.35 \times 10^{-4}\Big(\frac{\beta}{H}\Big)^{-1}v_w \Big(\frac{\epsilon \kappa_v \alpha}{1+\alpha}\Big)^{\frac{3}{2}}\Big(\frac{g_*}{100}\Big)^{-\frac{1}{3}}\frac{(\frac{f}{f_{turb}})^3\Big(1+\frac{f}{f_{\rm turb}}\Big)^{-\frac{11}{3}}}{\Big(1+\frac{8\pi f}{h_*}\Big)},
\eea
where $f_{\rm turb}$ represents the peak frequency contribution coming from the turbulence mechanism to the GW spectrum  and $\epsilon =0.1$:
\bea
f_{\rm turb} = 2.7 \times 10^{-5} Hz \Big(\frac{1}{v_w}\Big)\Big(\frac{\beta}{H}\Big)\Big(\frac{T_n}{100 \rm GeV}\Big)\Big(\frac{g_*}{100}\Big)^{\frac{1}{6}}.
\eea
where,
\bea
h_*=16.5 \times 10^{-6} Hz \Big(\frac{T_n}{100 \rm GeV}\Big)\Big(\frac{g_*}{100}\Big)^{\frac{1}{6}}.
\eea
For this analysis, the $\kappa_v$ supplied in \autoref{eq:4.22} is upgraded to the following expression \cite{Ellis:2018mja, Espinosa:2010hh}:
\bea
\kappa_v \simeq \Big[\frac{\alpha_{\infty}}{0.73+0.083\sqrt{\alpha_{\infty}}+\alpha_{\infty}}\Big] .
\eea

\begin{table}[h]
	\begin{center}
		\begin{tabular}{|c|cccccc|}\hline
			& $i$ & pattern & $T_i$[GeV]  & order &  $\alpha$ & $\beta/H_n$  \\ \hline
			& $T_c$ & 2 & 1377.9 & 2  &&\\
			BP1 & $T_c$ & 2 & 127.69  & 1   &&\\ 
			& $T_n$ & 2 & 127.68 & 1 &  5.36 & 1532.0\\ \hline
			BP2 & $T_c$ & 1 & 135.10  & 1   &&\\ 
			& $T_n$ & 1 & 135.06 & 1 &  3.42 & 319.14\\ \hline
			BP3 & $T_c$ & 1 & 130.81  & 1   &&\\ 
			& $T_n$ & 1 & 130.80 & 1 &  4.90 & 205.49\\ \hline
			BP4  & $T_c$ & 1 & 137.78  & 1   &&\\ 
			& $T_n$ & 1 & 137.77 & 1 &  3.24 &  360.99\\  \hline
			BP5 & $T_c$ & 1 & 137.22  & 1   &&\\ 
			& $T_n$ & 1 &137.21 & 1 &  3.38 & 324.24\\ 
			\hline
		\end{tabular} 
	\end{center}
	\caption{Phase transition strength  for the allowed benchmark points given in \autoref{tab:int1}-\autoref{tab:int2}. The critical temperature $T_c$ and nucleation temperature $T_n$ are displayed for each benchmark point, together with the "order" (first- and second-order phase transition) and "pattern" 1 or 2 (one- or two-step phase transition) for each. The corresponding bare mass parameters for the scalar triplets are 500 GeV for this analysis.}	\label{tab:table2}
\end{table}
The $\tt{CosmoTransition}$ \cite{Wainwright} package implements the effective potential in \autoref{eq:4.7} for calculating the pertinent parameters required for the computation of frequencies of the GWs. The variation of the minima of the potential with the 

\begin{figure}[h]
	\begin{center}
		\mbox{\subfigure[ ]{\includegraphics[width=0.5\linewidth,angle=-0]{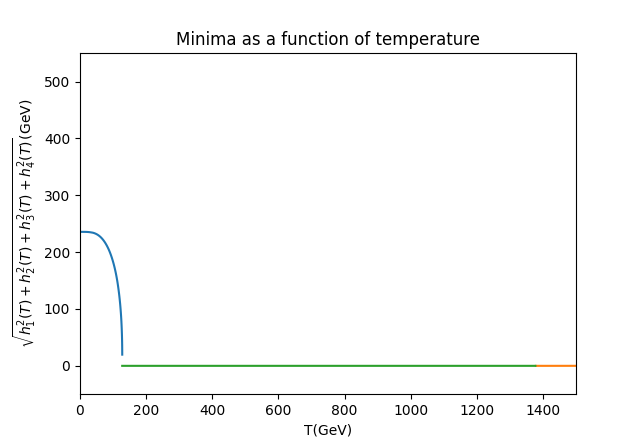}\label{f11a}}
			\subfigure[BP2]{\includegraphics[width=0.5\linewidth,angle=-0]{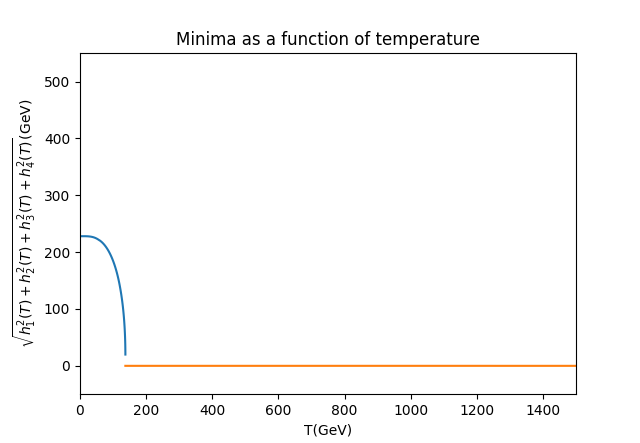}\label{f12a}}}
		\mbox{\subfigure[ ]{\includegraphics[width=0.5\linewidth,angle=-0]{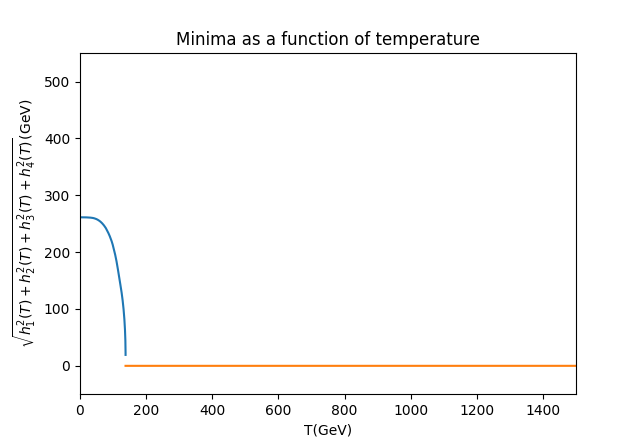}\label{f13a}}
			\subfigure[BP4]{\includegraphics[width=0.5\linewidth,angle=-0]{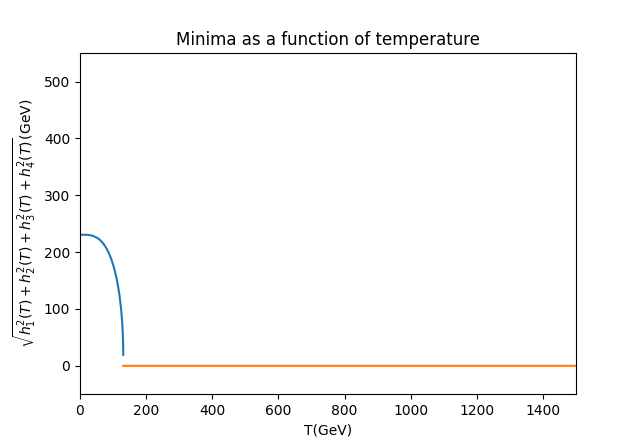}\label{f14a}}}
		\mbox{\subfigure[ ]{\includegraphics[width=0.5\linewidth,angle=-0]{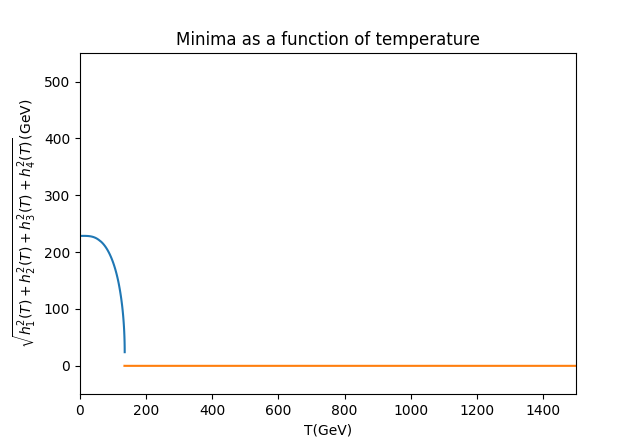}\label{f15a}}
		}
		\caption{Minima of the potential as a function of temperature in GeV, where $h_1$ is the background field for the SM Higgs doublet and $h_2, h_3$ and $h_4$ are the background fields for the Higgs triplets.}\label{fig:minima}
	\end{center}
\end{figure}
\begin{figure}[h]
	\begin{center}
		\mbox{\subfigure[ 500 GeV ]{\includegraphics[width=0.5\linewidth,angle=-0]{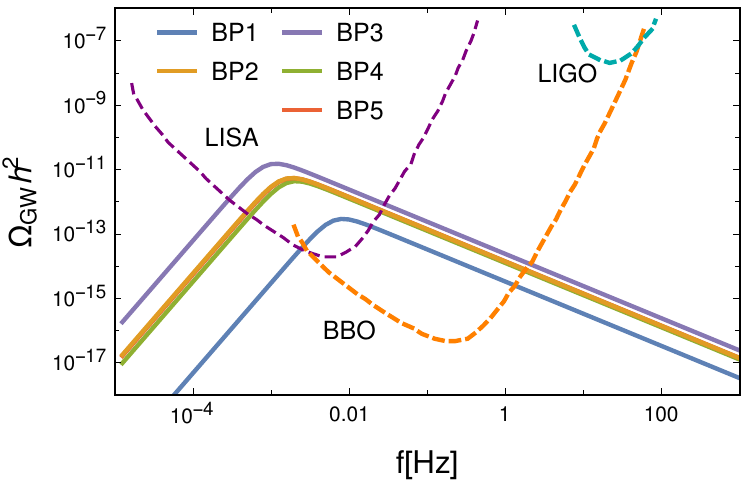}\label{f1a}}
			\subfigure[1 TeV]{\includegraphics[width=0.5\linewidth,angle=-0]{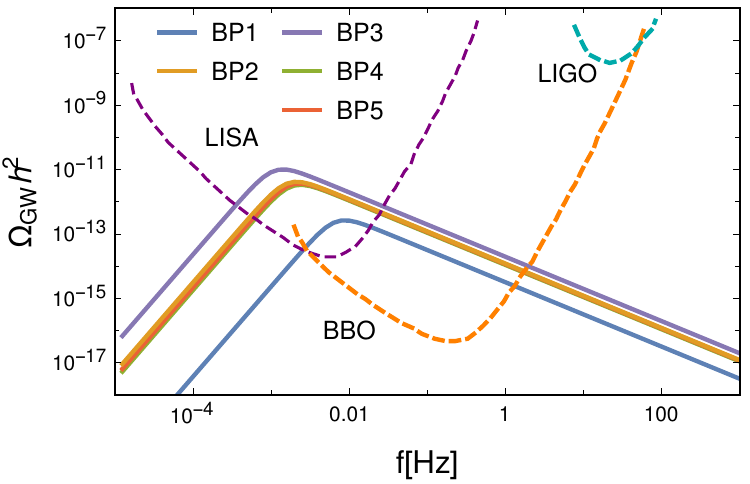}\label{f2a}}}
		\caption{GW signatures for the benchmark points permitted by strongly first-order phase transition and vacuum stability at the Planck scale. For BP2-BP5, the gravitational wave intensity peaks at about $10^{-3}$ Hz, whereas for BP1, it peaks at about $0.01$ Hz. For every benchmark point, the GWs intensity is within the observable frequency range of the BBO and LISA studies. For every BP, the GWs intensity is less intense for the corresponding LIGO detectable range.}\label{fig:gw}
	\end{center}
\end{figure}

temperature is plotted in \autoref{fig:minima} for BP1-BP5 from \autoref{tab:int1}- \autoref{tab:int2}. Each color denotes a different phase and discontinuity denotes the change of phase at a particular temperature. For BP1, the phase transition occurs in two steps; first around temperature $T_c =1377.9$ GeV which is second order, and later, second transition occurs around $T_c=127.69$ GeV which is first-order. At BP1, the electroweak phase transition proceeds through an intermediate phase, resulting in a two-step pattern. The first transition occurs between two nearby local minima of the finite-temperature effective potential, which are separated by a relatively shallow barrier. Because the vacuum expectation values involved in this step are small and the free-energy difference between the competing minima is limited, the corresponding change in the order parameter is not visually prominent in \autoref{fig:minima}(a).
	To verify that this transition is not a numerical artifact, we have carefully tracked the temperature evolution of all relevant field directions and explicitly monitored the emergence and disappearance of local minima in the full multi-field potential. We also confirmed the existence of two distinct critical temperatures by checking the degeneracy of the minima and ensuring the numerical stability of the solution under variations of the step size and convergence criteria used in the minimization procedure.
	The second transition, which corresponds to the onset of electroweak symmetry breaking, exhibits a much stronger barrier, and therefore dominates the phenomenological implications, such as the strength of the phase transition and the Gravitational-wave signal, and it is a genuine feature of the thermal potential at BP1 rather than a numerical artifact.\\
	The noticeably different values of the latent heat parameter $\alpha$ and the inverse duration parameter, $\beta/H_n$ at BP1 are indeed closely related to the multi-step nature of the electroweak phase transition in this benchmark scenario.
		At BP1, the phase transition proceeds through an intermediate phase before reaching the final electroweak-broken vacuum. As a result, the thermodynamic quantities relevant for Gravitational-wave production are dominantly determined by the second transition, which is responsible for the actual electroweak symmetry breaking. In this case, part of the vacuum energy is already released during the first, milder transition, reducing the amount of energy available at the second step. This leads to a modified latent heat, and hence, a different value of $\alpha$ compared to the single-step transitions realized at the other benchmark points.
		Similarly, the parameter $\beta/H_n$, which characterizes the time scale of the transition, is sensitive to the shape of the potential barrier and the temperature dependence of the nucleation rate. In a two-step transition, the second step typically occurs in a background that has already undergone partial symmetry breaking, resulting in a flatter effective potential near the nucleation temperature. This alters the temperature derivative of the Euclidean action and consequently yields a value of $\beta/H_n$ that differs significantly from those in single-step transitions.
		Therefore, the distinctive values of $\alpha$ and $\beta/H_n$ at BP1 arise from the redistribution of vacuum energy and the altered nucleation dynamics inherent to multi-step electroweak phase transitions. This highlights the sensitivity of Gravitational-wave observables to the detailed structure of the thermal history.
	 For BP2-BP5, the phase transition occurs via one-step and it is first order. The relevant parameters for the GW signatures, i.e., $T_c$, $T_n$, $\alpha$ and $\beta/H_n$ are given in \autoref{tab:table2}-\autoref{tab:table3} for two different values of bare mass parameters of the scalar triplet, i.e., 500 GeV and 1 TeV, respectively. The strength of the phase transition and the corresponding GW intensity is more or less same for both values of the bare mass parameters. The intensity of GWs with the frequency in Hertz is plotted in \autoref{fig:gw} for all the BPs for both 500 GeV and 1 TeV mass range of bare mass parameters of the scalar triplets.


\begin{table}[h]
	\begin{center}
		\begin{tabular}{|c|cccccc|}\hline
			& $i$ & pattern & $T_i$[GeV]  & order &  $\alpha$ & $\beta/H_n$  \\ \hline
			& $T_c$ & 2 & 1397.1 & 2  &&\\
			BP1 & $T_c$ & 2 & 129.45  & 1   &&\\ 
			& $T_n$ & 2 & 129.23 & 1 &  5.16 & 1601.03\\ \hline
			BP2 & $T_c$ & 1 & 137.15  & 1   &&\\ 
			& $T_n$ & 1 & 137.09 & 1 &  3.19 & 360.07\\ \hline
			BP3 & $T_c$ & 1 & 132.67  & 1   &&\\ 
			& $T_n$ & 1 & 132.53 & 1 &  4.68 & 250.17\\ \hline
			BP4  & $T_c$ & 1 & 139.41  & 1   &&\\ 
			& $T_n$ & 1 & 139.07 & 1 &  3.05 &  405.12\\  \hline
			BP5 & $T_c$ & 1 & 139.10  & 1   &&\\ 
			& $T_n$ & 1 &138.98 & 1 &  3.25 & 389.45\\ 
			\hline
		\end{tabular} 
	\end{center}
	\caption{Phase transition strength  for the allowed benchmark points given in \autoref{tab:int1}-\autoref{tab:int2}. The critical temperature $T_c$ and nucleation temperature $T_n$ are displayed for each benchmark point, together with the "order" (first- and second-order phase transition) and "pattern" 1 or 2 (one- or two-step phase transition) for each. The corresponding bare mass parameters for the scalar triplets are 1 TeV for this analysis.}	\label{tab:table3}
\end{table}

In order to put our benchmark points into context with other extended Higgs sectors, we briefly compare the sizes of the quartic couplings relevant for a strongly first-order electroweak phase transition (SFOEWPT) in the literature.
	For the Type-II seesaw model, a detailed study of the electroweak phase transition by \cite{Zhou:2022mlz}, finds that a SFOEWPT generically prefers positive Higgs portal couplings between the triplet and the SM Higgs and a relatively light triplet with mass below 
	$\sim$ 550 GeV. Although this work does not list exact coupling values in the abstract, the preference for positive and appreciable portal couplings implies that the corresponding quartic interactions in the Type-II seesaw potential are typically non-negligible and of moderate size in the parameter regions that support a strong EWPT. 
	By contrast, numerous studies of SFOEWPT in two-Higgs-doublet models (2HDM) indicate that achieving a strong transition tends to require order-one quartic couplings in the scalar potential. For example, parameter scans in Type-II 2HDM frameworks that enforce SFOEWPT find that the mass splittings and self-interactions among the additional Higgs fields, and hence the associated quartic coefficients in the potential, must be sizeable (typically $\lambda_i \sim \mathcal{O}(0.5 - 1.0)$ or above) to generate a sufficiently deep thermal potential barrier. This trend is seen in systematic investigations of SFOEWPT in Type-II 2HDMs, where benchmark scenarios with successful transitions are characterized by large inter-doublet quartic interactions. 
	In comparison, the quartic couplings at our benchmark points (\autoref{tab:int1}- \autoref{tab:int2}) are noticeably smaller than these typical values, yet we still find strongly first-order transitions. The key reason is that our model contains multiple interacting scalar triplets with Higgs portal and inter-triplet interactions. The combined finite-temperature contributions from these additional degrees of freedom enhance the thermal barrier between phases even when individual quartic couplings are modest. This multi-scalar enhancement mechanism means that a strong transition can be realized with smaller individual couplings than would be required in simpler two-scalar extensions such as the 2HDM or minimal Type-II seesaw.

The GW signatures for the benchmark points permitted by strongly first-order phase transition and vacuum stability at the Planck scale are displayed in \autoref{fig:gw}. In the present work, $v_w$ is not calculated dynamically from the microphysics of bubble expansion. Instead, it is taken as a fixed input parameter, which is a standard practice in many studies of first-order electroweak phase transitions. Specifically, for all benchmark points considered in this paper, we use $v_w =0.6$, corresponding to a moderately relativistic wall velocity. This choice lies between slow deflagration ( $v_w \lesssim 0.3$) and ultra-relativistic runaway scenarios (
	$v_w \rightarrow 1$) and, is commonly adopted as a representative value in phenomenological studies of Gravitational-wave production from electroweak phase transitions.
	The bubble wall velocity plays a crucial role in determining the strength and shape of the Gravitational-wave spectrum. The overall amplitude of the spectrum and the efficiency of energy transfer from the vacuum to the plasma are sensitive to $v_w$. In particular:
		A smaller wall velocity ($v_w \lesssim 0.3$) corresponds to slow-moving bubbles, leading to a larger fraction of the released energy being transferred to the plasma rather than directly into the scalar field, which modifies the relative contributions of the sound-wave and turbulence components.
	A larger wall velocity ($v_w \rightarrow 1$) corresponds to ultra-relativistic runaway bubbles, enhancing the scalar field contribution and shifting the peak frequency.
	Because of these theoretical uncertainties, we also checked that varying $v_w$ within a reasonable range (
	$0.3 \lesssim v_w \lesssim 0.9$) does not qualitatively change our conclusions. While the peak amplitude and peak frequency of the Gravitational-wave signal vary, the overall trends across different benchmark points and the detectability prospects with planned experiments such as LISA, LIGO, and BBO remain robust. A precise calculation of the bubble wall velocity in this multi-triplet scalar setup would require a full treatment of the friction and plasma dynamics.
	 
	For BP2-BP5, the GW intensity lies in the detectable frequency range of LISA and BBO with peak around $10^{-3}$ Hz. For BP1 with two-step transition, the GW intensity still lies in the detectable frequency range of LISA and BBO with peak around $0.01$ Hz. The Gravitational-wave sensitivity curves adopted in our analysis are taken from well-established references in the literature and correspond to the projected sensitivities of future and current experiments. In particular, the LISA sensitivity curve is based on the official LISA proposal and subsequent updates, as presented in Laser Interferometer Space Antenna (LISA), \cite{LISA:2017pwj}, the BBO sensitivity curve follows the standard projections discussed in \cite{Cutler:2005qq}, and the LIGO sensitivity curve corresponds to the design sensitivity of Advanced LIGO, as summarized in \cite{LIGOScientific:2014pky}.
 This makes the comparison between our predicted Gravitational-wave spectra and experimental sensitivities fully transparent. The frequency ranges for all the BPs lies in the detectable frequency range of LIGO but the corresponding GW intensity is lower compared to the detectable range.

\section{Experimental constraints}
In the present work, our primary focus is on the theoretical consistency of the extended scalar sector and its impact on the electroweak phase transition. Nevertheless, the benchmark points considered are chosen to be compatible with the existing experimental constraints.
Concerning direct searches, the additional scalar triplet states at the benchmark points have masses at or above the TeV scale. As a result, current LHC searches for doubly and singly charged scalars, which mainly constrain lighter states or scenarios with large production cross sections, do not exclude the benchmark configurations considered in this work.
Regarding electroweak precision observables, in particular the oblique parameters $S, T,$ and $U$, the dominant contributions arise from mass splittings among the scalar components. At the benchmark points, these splittings are kept moderate, ensuring that the resulting corrections remain within the experimentally allowed ranges. Moreover, the approximate custodial symmetry preserved in the scalar spectrum further suppresses potentially dangerous contributions to the $T$ parameter.
We emphasize that the purpose of the benchmark points is to illustrate the viability of a strong first-order electroweak phase transition within a theoretically consistent and experimentally allowed region of parameter space. A comprehensive collider-level analysis of the full parameter space would be an interesting extension of the present study but lies beyond its scope.
The benchmark points (BPs) considered here also have testable implications at current and future experiments.
	\subsection{Direct and indirect searches:}
	At the benchmark points, the extended scalar sector predicts additional neutral and charged scalar states with masses in the few-hundred-GeV to TeV range. Such states can be probed at the LHC and future colliders through direct searches in channels involving gauge boson pairs, multilepton final states, and, for triplet-like scalars, same-sign dilepton signatures. Indirect constraints from electroweak precision observables, particularly the oblique parameters, can be satisfied due to the relatively small mass splittings among the scalar states at the benchmark points, consistent with existing bounds.\\
		\subsection{Higgs trilinear coupling:}
A strongly first-order EWPT is known to be closely correlated with a deviation of the Higgs trilinear self-coupling from its Standard Model (SM) value, since both originate from the shape of the scalar potential and the presence of a potential barrier at finite temperature. In many extended Higgs sectors, such as the two-Higgs-doublet model or the Type-II seesaw model, achieving a strong first-order EWPT typically requires sizable quartic couplings, leading to deviations in the Higgs trilinear coupling of order $\Delta \lambda_{hhh}/\lambda^{SM}_{hhh} \sim \mathcal{O}(30\%- 100\%)$, as shown for example in \cite{Grojean:2004xa, Kanemura:2004mg, Chiang:2017nmu}. In contrast, in our model the strongly first-order EWPT is realized through the collective effect of multiple scalar degrees of freedom, rather than large individual quartic couplings. As a result, the quartic couplings at the benchmark points are comparatively moderate, and we expect the deviation in the Higgs trilinear coupling to be smaller but still potentially observable, typically at the level of $\Delta \lambda_{hhh}/\lambda^{SM}_{hhh} \sim \mathcal{O}(10\%- 30\%)$, which is comparable to what has been found in multi-scalar extensions with similar dynamics.
	Such deviations are within the projected sensitivity of future Higgs facilities. For instance, the HL-LHC is expected to probe the Higgs trilinear coupling at the $\sim 50 \%$ level, while future lepton colliders such as the ILC, FCC-ee, and CLIC can reach sensitivities of $\sim 10 \% - 20 \%$ (see, e.g., Refs. \cite{Cepeda:2019klc,deBlas:2019rxi}).\\
		\subsection{Scope of the present work:}
	A precise numerical determination of the Higgs trilinear coupling in our model would require a dedicated zero-temperature effective potential analysis, including loop corrections and renormalization-scheme dependence, which goes beyond the scope of the present study. Nevertheless, the benchmark points considered here clearly suggest a correlated signal: a strong EWPT accompanied by a potentially measurable deviation in the Higgs self-coupling and complementary collider signatures of the new scalar states.
\section{Conclusion}
In this work we have extended the Type-II seesaw model with
$U(1)_{L_{\mu}-L_{\tau}}$ symmetry where we require at least
three triplets to generate a realistic neutrino mass matrix which
has an inbuilt structure of two-zero texture. In this multi-fields
scalar sector we have explored the vacuum stability and the pertubativity
of the scalar potential. We have found that these triplets provide enough
positive contribution to ameliorate the Higgs vacuum instability issue and
the stability is achieved till Planck scale using two-loop $\beta-$ functions. Moreover, we
have noticed that the two-loop perturbativity is satisfied till $10^{12}$ GeV. We agree that the appearance of a Landau pole at a scale of 
	$\mu \sim 10^{12}$ GeV implies that the present model should be regarded as an effective field theory valid up to this scale, and that conclusions drawn beyond it are not reliable.
	Our statement regarding the improvement of the Higgs vacuum stability should therefore be understood in this effective-field-theory sense. In the Standard Model, the Higgs quartic coupling turns negative at an intermediate scale $\mu \sim 10^9 - 10^{11}$
	GeV, signaling the onset of vacuum metastability well below the Planck scale. In contrast, in the present model the additional scalar triplets provide positive loop contributions to the running of the Higgs quartic coupling, which prevent it from becoming negative throughout the entire range where the theory remains perturbative and under control, namely up to $\mu \sim 10^{12}$
	GeV.
	From this perspective, the introduction of the triplet scalars genuinely improves the vacuum stability problem within the domain of validity of the theory: the electroweak vacuum remains absolutely stable up to the cutoff scale set by perturbativity. While it is true that one cannot make definitive statements about stability above the Landau pole without specifying a UV completion, the absence of an instability below that scale represents a meaningful improvement compared to the Standard Model, where the instability typically arises at lower energies.
Next, we have considered this allowed parameter space from the vacuum stability
and perturbativity for the study of electroweak phase transition. It has been
found that the three triplets contribute significantly enough to the cubic term crucial
for generating the barrier between the symmetric phase and broken phase. As a result, the
phase transition comes out to be a strongly first-order till TeV scale mass range, i.e.,
until the degrees of freedom are heavy enough to decouple from the thermal bath.
Finally, we have also studied the Gravitational wave signatures for two different
values of the bare mass parameters of the triplets, i.e., 500 GeV and 1 TeV, respectively. The Gravitational wave signatures
are almost same for both the values of bare mass parameters of the scalar triplets, and lie in
the detectable frequency range of LISA and BBO experiments. 

\section{Acknowledgement}
This work was supported by the National Research
Foundation of Korea(NRF) grant funded by the Korea
government (MSIT) (RS-2024-00340153)(S.C.P.). This research was supported by an appointment to the YST Program at the APCTP through the Science and Technology Promotion Fund and Lottery Fund of the Korean Government. This was also supported by the Korean Local Governments - Gyeongsangbuk-do Province and Pohang city (S.J.).

\appendix
\section{Two-loop $\beta$-functions for dimensionless couplings} \label{betaf1}
\subsection{Scalar Quartic Couplings}\label{A1}
\footnotesize{
	\begingroup
	\allowdisplaybreaks

	\endgroup

	\bibliography{References}
	\bibliographystyle{Ref}
\end{document}